\documentstyle[own_sngl]{article}
\newcommand{\oversim}[2]{\protect{\mbox{\lower0.5ex\vbox{%
   \baselineskip=0pt\lineskip=0.2ex
   \ialign{$\mathsurround=0pt #1\hfil##\hfil$\crcr#2\crcr\sim\crcr}}}}} 
\newcommand{\simgreat}{\mbox{$\,\mathrel{\mathpalette\oversim>}\,$}} 
\newcommand{\simless} {\mbox{$\,\mathrel{\mathpalette\oversim<}\,$}} 
\slugcomment{{\em {\bf NewA, accepted: 26th June 1999}}}
\lefthead{Kroupa, Petr, McCaughrean}
\righthead{Binaries in Trapezium Cluster}
\begin{document}
%
\title {Binary stars in young clusters:
models versus observations of the \\Trapezium Cluster}
\author {Pavel Kroupa$^1$, Monika G. Petr$^{2,}$\footnote[4]{present
address: European Southern Observatory, Casilla 19001, Santiago 19,
Chile}, and Mark J. McCaughrean$^3$\\
\medskip
\small{$^1$Institut f{\"u}r Theoretische Astrophysik, Universit{\"a}t
Heidelberg, Tiergartenstr. 15, D-69121 Heidelberg, Germany\\
e-mail: pavel@ita.uni-heidelberg.de\\
$^2$Max-Planck-Institut f\"ur Astronomie, K\"onigstuhl~17, 
D-69117 Heidelberg\\ e-mail: mpetr@eso.org\\
$^3$ Astrophysikalisches Institut Potsdam, An der Sternwarte 16,
D-14482 Potsdam, Germany\\e-mail: mjm@aip.de\\
}} 
\begin{abstract}
\noindent 
The frequency of low-mass pre-main sequence binary systems is
significantly lower in the Trapezium Cluster than in Taurus-Auriga.
We investigate if this difference can be explained through stellar
encounters in dense clusters.  To this effect, a range of possible
models of the well observed Trapezium Cluster are calculated using
Aarseth's direct N-body code, which treats binaries accurately.  The
results are confronted with observational constraints. The range of
models include clusters in virial equilibrium, expanding clusters as a
result of instantaneous mass loss, as well as collapsing clusters.  In
all cases the primordial binary proportion is larger than 50~per cent,
with initial period distributions as observed in Taurus-Auriga and the
Galactic field.

It is found that the expanding model, with an initial binary
population as in the Galactic field, is most consistent with the
observational constraints. This raises the possibility that the
primordial group of OB stars may have expelled the cluster gas roughly
50~000~yr ago. The cluster's bulk expansion rate is thus a key
observable that needs to be determined.  The other models demonstrate
that the rapidly decreasing binary proportion, its radial dependence
and the form of the period distribution, together with structural and
kinematical data, are very useful diagnostics on the present and past
dynamical state of a young cluster. In particular, kinematical cooling
from the disruption of wide binaries is seen for the first time.
\end{abstract}


\section{INTRODUCTION} 
\label{sec:intro}

\noindent
To develop an improved understanding of star formation it is very
useful to empirically map variations of the binary-star properties and
the initial mass function with environment, such as metallicity,
density and temperature of the star-forming cloud.  However, a naive
interpretation of observational data may result in misleading
conclusions. For instance, a comparison of binary proportions in
different stellar populations, without taking into account the
dynamical history of the population, is ill-fated, because the
distribution of orbital elements of a population carries a memory of
the dynamical history (Kroupa 1998). One purpose of this paper is to
further illustrate, on the one hand this danger, and on the other
hand, the richness in understanding that can be gained through
stellar-dynamical studies of young stellar systems.

The Trapezium Cluster in the Orion star forming region is a young,
populous stellar cluster, which contains a negligible amount
($<10\,M_\odot$) of residual gas (Wilson et al. 1997; O'Dell 1999,
private communication) and for which a large amount of observational
data has been collected over decades.  Thus, it is an excellent target
for investigating the early dynamical evolution of a young cluster
with a primordial binary proportion at least as large as in the
Galactic field.  The cluster's age is estimated to be younger than
5~Myr (Prosser et al. 1994; Hillenbrand 1997), and probably
$\simless10^6$~yr. The central stellar number density is very high,
with the mean inter-stellar separation being about 6000~AU
(McCaughrean \& Stauffer 1994).  The Trapezium Cluster is a
well-suited sample for the study of star formation in a dense stellar
environment, as opposed to the low-density young stellar groups in the
Taurus-Auriga star forming region (Gomez et al. 1993; Briceno et
al. 1998).

Variations in stellar properties may occur with star formation
environment.  For example, the observed pre-main sequence binary
population in the Trapezium Cluster is significantly different than in
Taurus-Auriga. In the latter approximately all stars are in binary
systems (K\"ohler \& Leinert 1998; Duchene 1999; Ghez et al. 1997),
while in the Trapezium Cluster the binary proportion is significantly
lower (Prosser et al. 1994; Petr 1998; Petr et al. 1998; but see also
Padgett, Strom \& Ghez 1997), being comparable to what is found in the
Galactic field (Duquennoy \& Mayor 1991).  This, however, does not
necessarily imply that most Galactic field stars stem from
Trapezium-Cluster type assemblages, there being evidence that the
proportion of long-period binary systems is significantly below the
Galactic field value (Scally, Clarke \& McCaughrean, 1999).  The
binary population may well decrease further in the Trapezium Cluster,
if it is a bound entity. And it may have been higher at birth.

A stellar system with a half-mass radius of $R_{0.5}=0.1-0.3$~pc,
containing $N=1000-2000$ stars, with a mass $M_{\rm
cl}=300-600\,M_\odot$, has a relaxation time $t_{\rm
relax}=0.6-4.2$~Myr, which is the time-scale over which the cluster
looses memory of its initial conditions.  These are characteristic
observed values for the Trapezium Cluster.  The three-dimensional
velocity dispersion in the Trapezium Cluster is $\sigma \approx
4.3$~km/s, the crossing time being $t_{\rm cross}=2-7\times10^4$~yr
(an overview of the dynamical evolution of young clusters can be found
in Bonnell \& Kroupa 1999).  A typical star may thus have crossed the
central region many times, given that the age of the cluster is
$\simless1$~Myr.  This implies that collisions between stellar systems
will have been frequent. It also implies that the birth structure of
the cluster will have been eradicated.  Indications that this may be
the case comes from the apparent lack of sub-structure (Bate, Clarke
\& McCaughrean 1998).

The Trapezium Cluster may thus be just old enough to perhaps have
allowed an initial Taurus-Auriga binary population to have dynamically
evolved to the observed reduced value. Alternatively, it may well be
true that the lower binary proportion of the Trapezium Cluster as
compared to the Taurus-Auriga star forming region is due to different
star-forming conditions. Durisen \& Sterzik (1994) point out that
binary formation from fragmentation of collapsing and rotating clouds,
or from a gravitational instability of massive proto-stellar disks, is
more likely in low-temperature clouds.

That the binary proportion is eroded efficiently in a dense cluster
has been shown by Kroupa (1995a).  However, whether the observed
difference between the Trapezium Cluster and Taurus-Auriga can be
explained with this mechanism must be investigated by constructing
self-consistent models, that more closely match the properties of the
Trapezium Cluster.  If such a study shows that encounters are not
efficient enough for a wide variety of models of the Trapezium
Cluster, then the conclusion must be that the binary proportion
depends on star-forming conditions.  The aim of this paper is to begin
tackling this problem by comparing evolving model populations with the
constraints given by observations. 

In this paper we present the first steps towards more realistic fully
self-consistent models of embedded clusters in general, and of the
Trapezium Cluster in particular.  Section~\ref{sec:obs} summarises the
available observational constraints and Section~\ref{sec:models}
details the model assumptions. In Section~\ref{sec:code} we explain
the calculation and data-evaluation programmes. The results are
presented in Section~\ref{sec:res}, and Sections~\ref{sec:disc}
and~\ref{sec:conc} contain the discussion and conclusions,
respectively. An accompanying paper (Kroupa 1999) generalises these
results to different numbers of stars and cluster radii, with a view to
discussing the dynamical state of the Orion Nebula Cluster and the set
of viable initial conditions.

\section{OBSERVATIONAL CONSTRAINTS}
\label{sec:obs}
\noindent 
The aim of this investigation is to study the evolution of those
observables that allow the dynamical state of a young cluster and its
birth configuration to be constrained.  Such observables are the
velocity dispersion, the central stellar volume number density and the
radial dependence of the binary proportion.

\subsection{Total number of stars, cluster age, and central density}
\label{sec:numb_age}
\noindent
In an optical high-resolution study carried out with the Hubble Space
Telescope, 319 stars in a region with size $\sim 0.45\,{\rm pc} \times
0.45\,{\rm pc}$ were identified (Prosser et al. 1994).  In a near
infrared study covering a field of $\sim 5^{\prime} \times 5^{\prime}$
($\sim 0.65\,{\rm pc} \times 0.65\,{\rm pc}$) centred on the Trapezium
Cluster, McCaughrean et al. (1996) counted 700 systems (we call a
single as well as a binary star a system).  Binary stars with
separations below the 0.7~arcsec seeing limit were not resolved. This
corresponds to a resolution limit of $\approx314$~AU, or
log$_{10}P\approx6.3$, where $P$ is in days, for a system with a mass
of $1\,M_\odot$.

A high-spatial-resolution direct-imaging near-infrared study revealed
even more stars in the innermost centre of the Trapezium Cluster
(McCaughrean \& Stauffer 1994), and a remarkably high stellar density
of $\sim 5 \times 10^4$ stars per pc$^3$ was estimated.  McCaughrean
\& Stauffer (1994) calculated that 29~systems are within the central
spherical volume with a radius of $R=0.053$~pc.

From the analysis of the HR-Diagram, based on optical photometry and
spectroscopy, Hillenbrand (1997) derived a mean age for the whole
Orion Nebula Cluster less than 1~Myr, with an age spread of a
few~Myr. The age determination given by Prosser et al. (1994) is
limited to the stars associated with the Trapezium Cluster and led to
a medium age of $\sim 3 \times 10^5$~yr, with nearly no detectable age
spread.

\subsection{Velocity dispersion, $\sigma_{\rm op}$}
\label{sec:veld}
\noindent
The projected one-dimensional velocity dispersion in the observational
plane, within $r=0.41$~pc of the centre of the Trapezium Cluster, was
derived by Jones \& Walker (1988) to be $2.54\pm0.27$~km/s from their
proper-motion survey using photographic plates.  The data indicate no
anisotropy, but these authors, van Altena et al. (1988) and Tian et
al. (1996) note that the plate-reduction algorithms eliminate any
signature due to rotation and/or expansion or contraction. Thus, it is
presently unknown if the Trapezium Cluster is expanding or
contracting. The analysis of new absolute proper motions of a few
dozen stars in the inner region ($\simless 2.5$~pc) of the Orion
Nebula Cluster (ONC), of which the Trapezium Cluster is likely to be
merely the central part, however, suggests that there may be some
expanding motion (Frink, Kroupa \& R\"oser 1999). For the present
analysis we adopt the velocity dispersion obtained from the more
precise relative proper motions, which we take to be corrected for
radial bulk motions.

Photographic plates have a spatial resolution of typically 2~arcsec,
which corresponds to about 1000~AU in the Trapezium Cluster. Most
binary systems are therefore not resolved in these surveys.

\subsection{Binary proportion}
\label{sec:binf}
\noindent
Observations using the speckle holography technique in the core of the
Trapezium Cluster by Petr et al. (1998) resolve binary systems with
separations in the range $d_1=63$~AU to $d_2=225$~AU.  Of the 42
systems that appear projected within the central radius, $r=0.041$~pc,
six are OB stars and four are binary systems within the studied range
of separations.  The apparent binary proportion of the entire sample
is $f_{\rm app}=0.095\pm0.05$ (equation~\ref{eqn:binfapp} below), and
the binary proportion for low-mass ($m<1.5\,M_\odot$) stars is $f_{\rm
app,lms}=0.06\pm0.04$.  The subscript ``app'' means that the observed
$f$ is the apparent binary proportion that an observer deduces from
projected star positions within some range of separations to which the
observational apparatus is sensitive.  The binary proportion of
low-mass stars within the central radius $r=0.25$~pc is found by
Prosser et al. (1994) using HST imaging to be $f_{\rm
app}=0.12\pm0.02$ in the separation range $d_1=26$~AU
to~$d_2=440$~AU. The orbital periods corresponding to the respective
separations are given in Fig.~\ref{fig:fappA}.

Both studies thus find a binary proportion, in the respective distance
ranges sampled, that is similar to the Galactic field value, implying
no measurable radial dependence of $f$ for $r<0.3$~pc (for details see
Prosser et al. 1994, and Petr et al. 1998). 

\section{MODEL CLUSTERS}
\label{sec:models}
\noindent
In this section we construct the model clusters. The appearance of the
Trapezium Cluster today does not reflect the initial conditions when
the stars decoupled dynamically from the gas, because the velocity
dispersion of 2.5~pc/Myr implies significant mixing or spreading
within 0.5--1~Myr. We therefore realize a variety of initial dynamical
states. These cover models in cold collapse through initially
virialised systems to clusters that expand.

\subsection{The stellar population}
\label{sec:stpop}
\noindent
The following parameters are chosen to specify possible different
model Trapezium Clusters at ``birth'', i.e. at the time when stellar
dynamics starts dominating over gas dynamics:
\begin{itemize}
\item $N = 1600$ stars (point masses);
\itemsep -1mm
\item The initial mass function (IMF) deduced from a careful analysis
of star-count data in the Galactic field (Kroupa, Tout \& Gilmore
1993) with power-law index $\alpha_1=1.3$ for $m\le 0.5\,M_\odot$;
this IMF is based on Scalo's (1986) IMF for stars with $m>1M_\odot$;
\itemsep -1mm
\item Lower and upper stellar mass limits of $m_{\rm
l}=0.08\,M_\odot$ and $m_{\rm u}=30\,M_\odot$, respectively;
\itemsep -1mm
\item A Plummer density distribution (Aarseth, Henon \& Wielen 1974);
\itemsep -1mm
\item The initial position and velocity vectors of the binary-star
centre-of-masses and single stars are independent of stellar mass;
\itemsep -1mm
\item The velocity distribution of the binary centre-of-masses and
single stars is isotropic.
\end{itemize}
Models in virial equilibrium (A1, A2), expanding models (B1, B2) and
two cases of a cold collapse (models~C1 and~C2) are computed.  Details
of the models are given in Table~\ref{table:models}, the entries of
which are described below and in the following sub-sections.  

The resulting cluster mass is $M_{\rm cl}=700\,M_\odot$.  The models
contain no gas mass. This is a safe assumption because Wilson et
al. (1997) and C.R.~O'Dell (private communication) estimate the mass
of gas to be only a few $M_\odot$ in the Trapezium Cluster.  The
choice of the initial number of stars is guided by the evidence
presented in Section~\ref{sec:numb_age}, remembering that not all
Trapezium Cluster stars will have been detected, as binary systems are
usually not resolved in lower spatial resolution imaging studies.  A
binary proportion of~60~per~cent implies $N=1120$ stars. However, we
choose $N=1600$ since rapid reduction of the number of systems in the
observed area ($0.65\times0.65$~pc$^2$) is to be expected owing to
expansion of the cluster and ejection of stars.  Severe computational
complications are avoided by assuming the stars have no extent.  The
assumption that the velocity and position vectors are not correlated
with the stellar mass contradicts the observations, which show that
the massive stars are concentrated in the cluster core. However, the
present assumption allows investigation of whether dynamical
mass-segregation in young binary-rich clusters can lead to the
observed mass-segregation on time-scales of 1~Myr or less. This will
be the subject of another paper. The initial placement of the massive
stars in the present models is also consistent with the empirical
finding that ultra-compact HII regions in areas of rich and active
star formation are usually distributed throughout the region over
spatial scales of about 1~pc (e.g. Megeath et al. 1996).

\subsection{Initial dynamical state}
\label{sec:initstate}
\noindent
The initial half-mass radius, $R_{0.5}$, is listed in column~2 of
Table~\ref{table:models}.  For models~A1 to~B2 it is based on the
results of McCaughrean \& Stauffer (1994).  Columns~3 and~4 list,
respectively, the number of systems and stars within the central
radius $R=0.053$~pc, and the central density ($\rho_{\rm
c}=3\,N/[4\,\pi\,(0.77\,R_{0.5})^3]$, where $0.77\,R_{0.5}$ is the
Plummer radius) is given in column~5.  The parameters for models~A1
and~A2 are a compromise in that they give a central density that is
larger by an order of magnitude than the observed value, and a
velocity dispersion that is somewhat smaller than the observed value
for virial equilibrium. The central density corresponds to an
inter-system spacing of~2200~AU.  In this case, the initial model
median relaxation and half-mass diameter crossing times are $t_{\rm
relax}=0.62$~Myr and $t_{\rm cross}=0.059$~Myr, respectively. The
central number density decreases and the half-mass radius increases
within about $10^6$~yr owing to mass segregation and associated
expansion, and heating through binary stars. The precise evolution
cannot be anticipated though, because it is not possible to easily
estimate the various channels of energy exchange that exist in such a
complex gravitational system, and since such realistic computations
have never been been performed heretofore.

Column~6 of Table~\ref{table:models} contains the initial virial
ratio, $Q=E_{\rm kin}/|E_{\rm pot}|$, where the nominator and
denominator are, respectively, the total kinetic and potential energy
of the whole cluster ($E_{\rm kin}=0.5\,\sigma^2M_{\rm cl}$ and
$E_{\rm pot}=-G\,M_{\rm cl}^2/R_{\rm G}$, where $M_{\rm cl}$ is the
cluster mass, $\sigma$ is the three-dimensional velocity dispersion,
$R_{\rm G}$ is a characteristic cluster radius, and $G$ is the
gravitational constant).  One possible state is virial equilibrium,
$Q_{\rm v}=0.5$.  However, the cluster may be expanding or contracting
(see Section~\ref{sec:veld}). To cover such possibilities, cluster
models in virial equilibrium, expanding ($Q_{\rm exp}=1.2$) and cold
collapse ($Q_{\rm col}=0.01$) are studied.

If the total cluster mass before mass loss is $M_{\rm tot}$, and if
the cluster is initially in virial equilibrium, then for a sudden mass
loss of amount $M_{\rm gas}$, the new 
\begin{equation}
Q_{\rm exp} 
        = {Q_{\rm v}\over 1-M_{\rm gas}/M_{\rm tot}}\\
        = {Q_{\rm v} \over \epsilon},
\end{equation}
where $\epsilon=M_{\rm cl}/M_{\rm tot}$ is the star-formation
efficiency.  Assuming the cluster instantly looses a mass $M_{\rm
gas}=0.58\,M_{\rm tot}$ ($\epsilon=0.42$), $Q_{\rm exp}=1.2$, and the
velocities are larger by a factor of $(Q_{\rm exp}/Q_{\rm
v})^{1/2}=\epsilon^{-1/2}=1.55$ than in a cluster in virial
equilibrium with a mass $M_{\rm cl}=M_{\rm tot} - M_{\rm
gas}=0.42\,M_{\rm tot}$.  Thus, in the case of the expanding models
(B1, B2), the initial velocities are increased by a factor
of~1.55. These models assume that the massive stars have driven out a
gas mass, $M_{\rm gas} = 967\,M_\odot$, immediately after they ``turn
on'', which is the time when the stellar-dynamical computation begins
($t=0$). This mass loss corresponds to a rather high star-formation
efficiency of~42 per~cent, which is nevertheless expected to lead to
an expanding, unbound association (Lada, Margulis \& Dearborn
1984). It is also in-line with the masses of outflows observed around
young OB stars which have short ($\simless10^4$~yr) durations
(Churchwell 1997). In such models it is possible that most of the
cluster existed in an embedded phase for a few $10^5$~yr before the
hypothetical gas expulsion event. This phase, however, is not modelled
here.

Models~C1 and~C2 are clusters which undergo cold collapse from two
different initial $R_{0.5}$. To achieve this, the initial velocities
of all binary centre-of-masses and single stars are multiplied by a
factor $(Q_{\rm col}/Q_{\rm v})^{1/2}=0.14$ (zero initial velocities
are problematical for the N-body programme and would also be
unrealistic since some clump--clump motions will always be present in
the proto-cluster).  Such models are interesting because dynamical
friction on dense gas clumps moving in an extended gas medium is
expected to lead to the contraction of a proto-cluster (Saiyadpour,
Deiss \& Kegel 1997; Ostriker 1999). Furthermore, relative
ammonia-clump--\-C$^{18}$O-core velocities in larger (by a factor of
2--3) C$^{18}$O cores are often observed to be significantly smaller
($\approx 0.1$~km/sec) than the observed velocity dispersion ($\approx
0.5$~km/s) of the core, which is thought to be close to the virial
value of the core (fig.~45 in Benson \& Myers 1989).  In this picture,
individual stellar systems remain trapped in the potential wells
locally dominated by the clumps that formed them, until most of the
gas is removed.  Clearly, relative clump--clump velocities need to be
measured for cores containing many clumps, to verify if this picture
is correct.  Finally, the proper motion measurements within the
Trapezium Cluster (Section~\ref{sec:veld}) cannot exclude a model in
which the cluster is collapsing, or is in the phase of violent
relaxation (Lynden-Bell 1967) after a cold collapse.

\subsection{Primordial binary systems}
\label{sec:prbins}
\begin{figure}
\plotfiddle{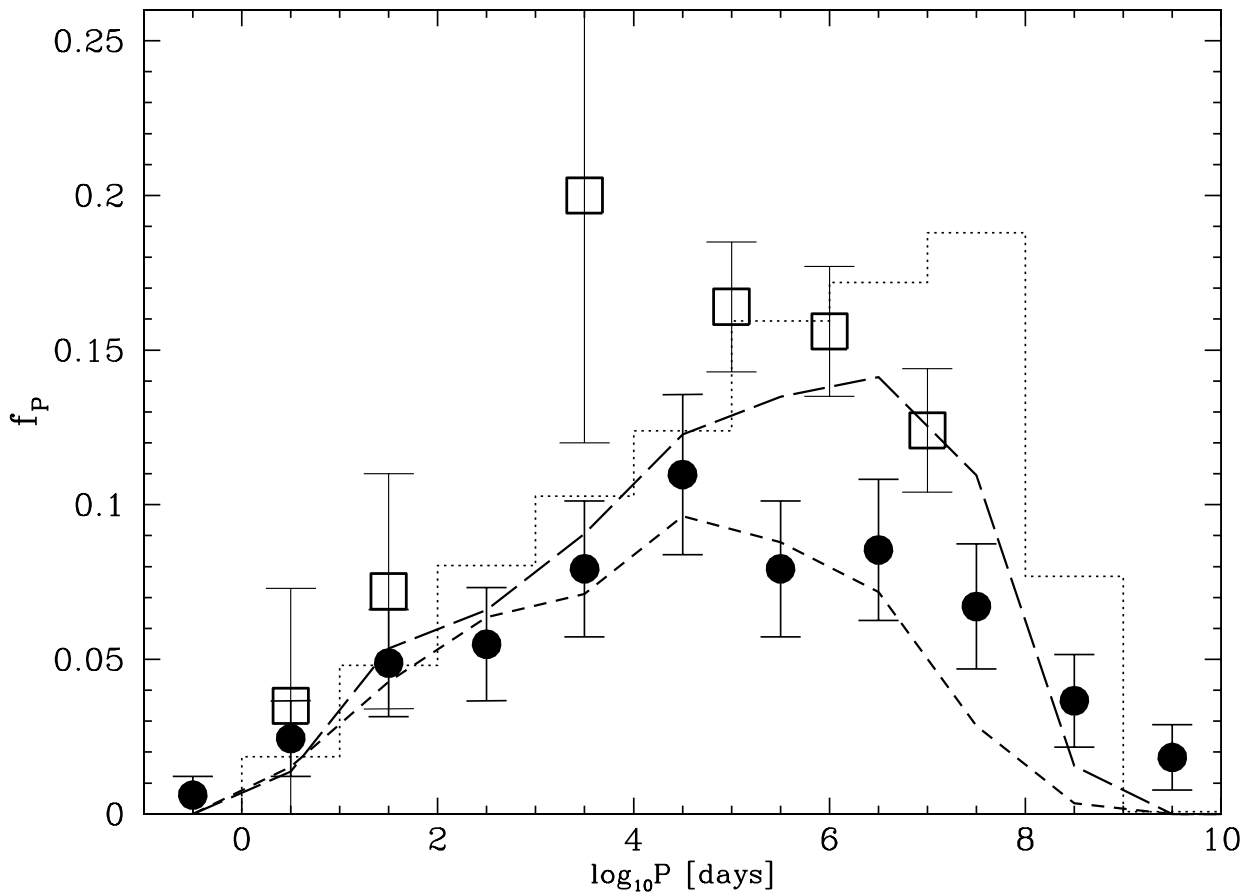}{15cm}{-90}{63}{63}{-250}{400}
\caption{Distribution of orbits, $f_{\rm P}$, for main sequence
multiple systems (solid dots, Duquennoy \& Mayor 1991) and pre-main
sequence systems in Taurus--Auriga (open squares; log$_{10}P>4$:
K\"ohler \& Leinert 1998, log$_{10}P=3.5$: Richichi et al. 1994,
log$_{10}P<2$: Mathieu 1994).  Main sequence G--, K-- and M--dwarf
binaries have essentially the same period distribution (fig.~1 in
Kroupa 1995a).  The observational data include triple systems, which
are counted as two orbits. Quadruple systems add three orbits. The
dotted histogram is the primordial period distribution from Kroupa
(1995b, fig.~7) with $f_{\rm tot}=1$. Crowding in Trapezium Cluster
models~A and~B changes this distribution to the initial long-dashed
one. Thus, a primordial $f_{\rm tot}=1$ becomes $f_{\rm tot}=0.77$
initially.  A primordial Gaussian log-period distribution with $f_{\rm
tot}=0.6$ that fits the solid dots, changes in the models to the
initial distribution shown as the short-dashed line, with $f_{\rm
tot}=0.48$. For model~A, the initial hard/soft binary boundary (Heggie
1975) is log$_{10}P_{\rm th}=5.4$, which is the orbital period for an
orbit with a circular velocity equal to the velocity dispersion in the
cluster (three-dimensional velocity dispersion
$\sigma=\sqrt{3}\,\sigma_{\rm op,c} = 3.46$~km/s,
Fig.~\ref{fig:d_vdA}). Systems with $P>P_{\rm th}$ are {\it soft},
whereas systems with $P<P_{\rm th}$ are {\it hard}.
\label{fig:init_P}}
\end{figure}

\noindent
For each model cluster different assumptions concerning the primordial
binary star population are made.  In Table~\ref{table:models},
column~7 lists the primordial total binary proportion,
\begin{equation}
f_{\rm tot} = {N_{\rm bin} \over N_{\rm bin}+N_{\rm sing}}, 
\label{eqn:binf}
\end{equation}
where $N_{\rm sing}$ and $N_{\rm bin}$ are the numbers of single-star
and bound-binary systems, respectively.  Column~8 contains the maximum
binary-star period.  The minimum period is $P_{\rm min}=1$~day in all
cases. The form of the primordial period distribution is defined in
column~9, where K2 refers to the eigenevolved (see below) birth
distribution, which is consistent with young systems in
Taurus--Auriga, and DM91 (Duquennoy \& Mayor 1991) refers to the
Gaussian log-period distribution approximating the distribution of
systems in the Galactic field. A $1\,M_\odot$ system has, with
log$_{10}P_{\rm max}=8.43$ (11.0) a semi-major axis of 8200~AU
($4.2\times10^5$~AU). The model period distributions are compared with
the observational data in Fig.~\ref{fig:init_P}. Note that the {\it
primordial} $f_{\rm tot}$ is immediately reduced to the {\it initial}
$f_{\rm tot}$ owing to disruptive crowding, which means through the
immediate non-dynamical ionization of long-period binaries owing to
their overlap in the high-density central region of the cluster.

With $f_{\rm tot}=1$, the assumption is made that the binary-star
properties do not vary with star-forming conditions, apart from the
effects of crowding, and that they are identical for the Trapezium
Cluster to what is observed in Taurus--Auriga.  With $f_{\rm
tot}=0.6$, the hypothesis is that the binary proportion is lower in
higher temperature star-forming clouds (Durisen \& Sterzik 1994).

Additionally, the following assumptions are made:
\begin{itemize}
\item The primordial mass-ratio distribution is obtained by random
pairing of the stars. 
\itemsep -1mm
\item The primordial eccentricity distribution is thermally relaxed.
\end{itemize}
The first assumption implies a mass-ratio distribution for G-dwarf
systems that increases steeply towards low-mass secondaries.  However,
in the Galactic field (Duquennoy \& Mayor 1991), the mass-ratio
distribution for G-dwarf binary systems contains fewer M~dwarf
companions than the initial model here. The assumption and
Galactic-field distribution are consistent with each other, because
dynamical interactions in clusters can explain the observed properties
of binary stars in the Galactic field, this being part of the {\it
inverse dynamical population synthesis} argument for a clustered
origin of most stars (Kroupa 1995a, 1995b).  The available
observational evidence on the mass-ratio distribution of very young
stars is also consistent with the assumption of random-pairing from
the IMF (Leinert et al. 1993).  

The second assumption is necessary, because dynamical interactions in
a Galactic cluster cannot evolve an arbitrary initial eccentricity
distribution to the thermal distribution observed for main sequence
binary systems with log$_{10}P\simgreat3$, in which the number of
systems increases linearly with the eccentricity.  The primordial
distribution is, however, evolved prior to the start of the $N$--body
integration to account for {\it pre-main sequence eigenevolution}
(i.e. the changes in orbital parameters owing to processes internal to
a binary system such as tidal circularisation during the pre-main
sequence phase). Only binary systems with log$_{10}P\simless3$ are
affected by this. Details can be found in Kroupa (1995b).

\section{THE N-BODY PROGRAMME AND DATA ANALYSIS} 
\label{sec:code}

\subsection{Aarseth's {\sc Nbody5}}
\noindent
\label{sec:nb5}
The computation of the dynamical evolution of star clusters is
expensive because the forces have to be calculated for each pair of
stars so that the CPU time scales as $N^2$. Binary systems that are
perturbed by neighbours, and thus need to be integrated with very small
time-steps, cause additional serious bottlenecks. An $N$--body
programme must be able to handle dynamical processes that have
time-scales ranging from days to many $10^8$~yr.  The state-of-the art
programmes for the dynamical modelling of star clusters on
commercially available personal computers have been developed by
Aarseth (1985, 1994, 1999).

For the present investigation, the successful and well tried {\sc
Nbody5} programme is used. It incorporates many special algorithms to
ensure computational speed and efficiency.  Each particle is advanced
with its own time-step, and the accelerations acting on it are updated
with different frequencies using different techniques, depending on
whether the perturbing stars are in a near-neighbour list or if the
forces stem from numerous distant cluster members. Binary, triple and
higher order systems are treated with the regularisation technique
which, with a special transformation of space-time coordinates,
eliminates the time-step singularity for closely interacting stars
(see Aarseth 1999 for the state-of-the art).

The version of {\sc Nbody5} employed here has been used earlier for
the calculation of the dynamical evolution of binary-rich clusters by
Kroupa (1995a,b,c). It has extra routines that allow the inclusion of
binary star populations with a variety of different initial period and
eccentricity distributions.  Position and velocity vectors of each
star are output at regular time intervals, and form the basic data
from which all binary star properties are derived in subsequent data
analysis.

While statistical results from $N$--body calculations correctly
describe the overall dynamical evolution, the velocity and position
vectors of any individual centre-of-mass particle (e.g. a star or
binary system) diverge exponentially from the true trajectory, through
the growth of errors in $N$--body computations. Goodman, Heggie \& Hut
(1993) discuss these issues in depth.  $N_{\rm run}=3$ computations
with different initial random number seeds are performed for each
model discussed here, giving in total 18~$N$--body calculations. All
quantities inferred from the models and presented here are averages
from three models.  The computations are performed for a time-span of
5~Myr only, which is the upper limit for the age of the Trapezium
Cluster from HR--diagram fitting (Hillenbrand 1997), although the mean
age is probably closer to 1~Myr.  Mass loss through stellar evolution
can thus be neglected.

\subsection{Data evaluation programme}
\label{sec:dateval}
\noindent
An elaborate data-reduction programme reads the output from {\sc
Nbody5} and calculates, among many quantities, the projected velocity
dispersion, the number of bound binary systems and their properties.
The observational plane is taken to be the $y-z$ plane, with the $z$
direction being perpendicular to the Galactic disk.  Throughout this
paper, $r$ refers to the projected radial distance from the density
maximum, whereas $R$ refers to the 3D distance.  The choice of the
projection plane is inconsequential for the present analysis, because
$r$ is always smaller than a few~pc.

The apparent binary proportion,
\begin{equation}
f_{\rm app} = {N_{\rm app,bin} \over N_{\rm app,bin}+N_{\rm app,sing}},
\label{eqn:binfapp}
\end{equation}
a hypothetical observer located at infinity sees in the projected
data, is calculated from the projected positions of all stars seen
within a central circle with radius $r$ in one model cluster at time
$t$. The number, $N_{\rm app,bin}$, of apparent stellar pairs with
separations in the distance range $d_1$ to $d_2$ are counted together
with the number of apparently single stars, $N_{\rm app,sing}$.
Binary systems with a separation $d<d_1$ are counted as single stars,
and binary systems with $d>d_2$ are counted as two single stars. This
is done for all stars within a central radius $r$.

For a comparison with proper motion data, the one-dimensional velocity
dispersion of centres-of-masses (i.e. unresolved stellar systems) in
the observational plane, $\sigma_{\rm op}$ is calculated within some
$r$.  For consistency with the observational data, $\sigma_{\rm op}$
is corrected for the bulk radial motion giving $\sigma_{\rm op,c}$. To
this end, the mean projected radial velocity, $\left< v_{\rm
r}\right>$, within $r$ is calculated. The corresponding radial vector
quantity is subtracted from each 2D velocity vector. In the
equilibrium models (A), $\left<v_{\rm r}\right>\approx0$, while for
the expanding models (B), $\left<v_{\rm r}\right> > 0$, and for the
cold collapse models (C), $\left<v_{\rm r}\right> < 0$, for $t>0$. The
bulk radial motion is zero at $t=0$ in the equilibrium and expanding
models, because the velocity vectors are initially isotropically
distributed.


\section{RESULTS}
\label{sec:res}
\noindent
In this section the results are described for each model (virial
equilibrium, expansion and cold collapse) in turn.  The overall
cluster evolution and the evolution of the binary star population, as
seen by a hypothetical observer, are discussed in separate
subsections. Throughout the discussion the binary-star orbital period
is used instead of the binding energy, which is $e_{\rm
b}=-P^{-2/3}\left[m_1m_2/\left(m_1+m_2\right)^{1/3}\right]G/2$, where
$m_1$ and $m_2$ are the component masses, and $G$ is the gravitational
constant in appropriate units.  A detailed theoretical treatment of
energy exchanges of binary systems with field stars is given by Heggie
(1975).

\subsection{Clusters in virial equilibrium -- A}
\label{sec:virequA}
\noindent

\subsubsection{Cluster evolution}
\label{sec:evolA}
\begin{figure}
\plotfiddle{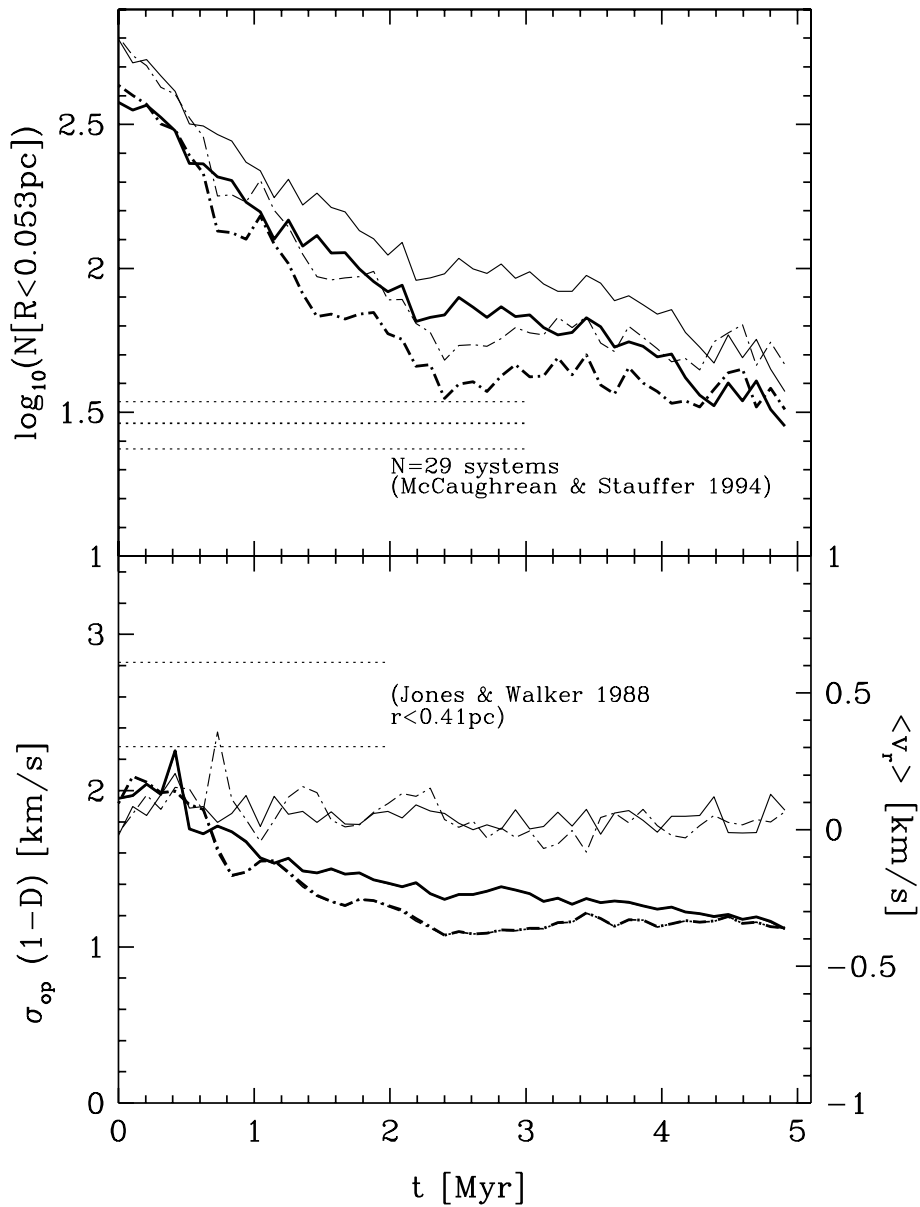}{15cm}{0}{64}{64}{-190}{-20}
\caption{Models~A1 and~A2.  In both panels, solid curves are for
primordial $f_{\rm tot}=1$ (model~A1), and dot-dashed curves are for
primordial $f_{\rm tot}=0.6$ (model~A2).  \underline{Upper panel}:
Time evolution of the number of stellar systems within a central
radius $R=0.053$~pc.  Thick curves assume no binary systems are
resolved, and thin curves count all stars.  The observational
constraint with the Poisson error range is indicated by the dotted
lines. \underline{Lower panel}: Time evolution of the one-dimensional
velocity dispersion in the observational plane of centre-of-masses
within a projected central radius $r=0.41$~pc ($\sigma_{\rm op}$,
thickest curves).  Thinnest curves are the projected mean radial
velocity within $r=0.41$~pc ($\left<v_{\rm r}\right>$, right
ordinate), while the curves with intermediate thickness show the
projected velocity dispersion corrected for the radial bulk motion,
$\sigma_{\rm op,c}$ (overlapping here with $\sigma_{\rm op}$), and are
the model quantities to be compared with the observational
constraints.  The observational one-sigma uncertainty range for
$\sigma_{\rm op,c}$ is indicated by the dotted lines. The initial
relaxation time for models~A1 and~A2 is $t_{\rm relax}=0.62$~Myr.
\label{fig:d_vdA}}
\end{figure}

\noindent 
The models are initially over-dense by an order of magnitude
(Fig.~\ref{fig:d_vdA}), but the central system density agrees with the
observational constraint after 2.3~Myr (model~A2) and 4.2~Myr
(model~A1). The cluster expands mostly through three- and four-body
interactions. In a cluster consisting initially of single stars, on
the other hand, the central number density would increase continuously
until core collapse (e.g. Heggie \& Aarseth 1992; Spurzem \& Takahashi
1995).  That the number of systems is consistently higher in model~A1
than in model~A2 for $t>0.5$~Myr is a result of enhanced binary
destruction in model~A1, which contains a larger number of binaries
with long periods.  The disruption of wide binaries cools cluster~A1,
which expands more slowly than cluster~A2, leading to a larger density
in the core.

The one-dimensional velocity dispersion (Fig.~\ref{fig:d_vdA}) is
similar to the observed value for $t\simless 0.5$~Myr, but then
decreases substantially. The velocity dispersion is significantly
smaller than the observational constraint when the central number
density agrees with the observations.  The higher density leads to a
higher velocity dispersion in model~A1.  This is a manifestation of
the negative specific heat capacity of gravitating systems: cooling
implying a hotter system (Lynden-Bell 1998; see also Bonnell \& Kroupa
1999).

In both models~A1 and~A2, the mean projected radial velocity,
$\left<v_{\rm r}\right> \approx +0.08$~km/s for $t\simless2$~Myr,
testifying to the expansion noted above. At later times $\left<v_{\rm
r}\right> \approx +0.05$~km/s, which corresponds to the smaller rate
of decrease of the central number density evident in the upper panel
of Fig.~\ref{fig:d_vdA} after~2.2~Myr. This is linked to the end of
the main destruction phase of long-period binary systems.  The small
radial bulk motion, $\left<v_{\rm r}\right>$, has a negligible effect
on the observed velocity dispersion ($\sigma_{\rm op,c}$).

The change in slope in the upper panel of Fig.~\ref{fig:d_vdA} at
$t\approx2.2$~Myr comes about because binary destruction moves the
truncation of the period distribution to decreasing periods and thus
towards the hard/soft binary boundary (see
Fig.~\ref{fig:init_P}). Meanwhile binary activity (i.e. the energy
exchange between the binary systems and the cluster field) expands the
cluster, causing a shift of the hard/soft boundary to longer
periods. Ultimately, at time $t_{\rm t}$ further binary destruction is
effectively halted, at which stage the hard/soft boundary has moved
beyond the truncation period of the period distribution, which in turn
implies a reduction of heating of the cluster field by hardening
binaries. The hard/soft, or {\it thermal}, boundary corresponds at
this stage to $P_{\rm th}=10^{6.0}$~days, taking the three-dimensional
velocity dispersion $\sigma=\sqrt{3}\,\sigma_{\rm op,c}=2.1$~km/s to be
the circular velocity of a system with a total mass of $1\,M_\odot$.
Thus, at $t>t_{\rm t}\approx2.2$~Myr the cluster retains an
approximately constant density, an equilibrium state between the
remaining binary population and the cluster field having been
established.

\subsubsection{The binary stars} 
\label{sec:binariesA}
\begin{figure}
\plotfiddle{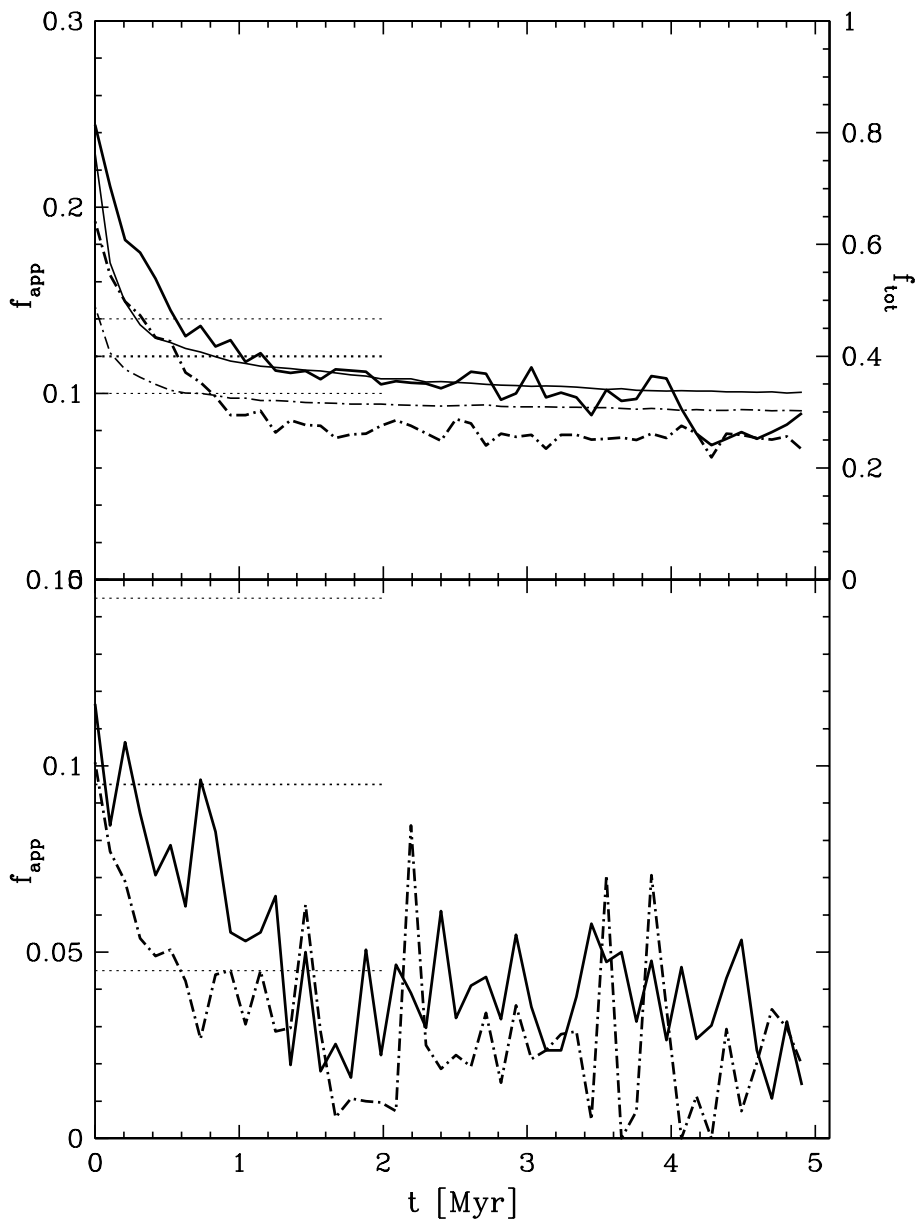}{15cm}{0}{62}{62}{-190}{-20}
\caption{Time evolution of the projected apparent binary proportion
(thick lines).  Solid line is for primordial $f_{\rm tot}=1$
(model~A1) and dot-dashed line is for primordial $f_{\rm tot}=0.6$
(model~A2). \underline{Upper panel}: Observational constraints on
$f_{\rm app}$ from Prosser et al. (1994, $r=0.249$~pc, $d_1=26$~AU,
$d_2=440$~AU, see Section~\ref{sec:dateval}) are shown as dotted
lines. Thin lines are the proportion of all bound binary systems,
$f_{\rm tot}$ (right ordinate).  \underline{Lower panel}:
Observational constraints from Petr et al. (1998, $r=0.04$~pc,
$d_1=63$~AU, $d_2=225$~AU) are shown as the horizontal lines. The
central dotted line is $f_{\rm app}$ for all systems in their sample,
which is to be compared with the theoretical results. Poisson
uncertainties are indicated by the upper and lower horizontal
lines. For comparison with Fig.~\ref{fig:per1}, a semi-major axis
$a=26$~AU corresponds to an orbital period in days of log$_{10}P=4.7$
for a $1\,M_\odot$ system. Similarly, $a=440$~AU corresponds to
log$_{10}P=6.5$, $a=63$~AU corresponds to log$_{10}P=5.3$, and
$a=225$~AU corresponds to log$_{10}P=6.1$. The initial crossing time
for models~A1 and~A2 is $t_{\rm cross}=0.06$~Myr.  \label{fig:fappA}}
\end{figure}

\noindent
The theoretical apparent projected binary star proportion, $f_{\rm
app}$, is compared with observational constraints in
Fig.~\ref{fig:fappA}.  Significant evolution of $f_{\rm app}$ occurs
within the first 1~Myr, owing to disruption of binaries and
diminishing number of chance projection pairs as the cluster expands.
However, $f_{\rm app}^{\rm model A1}>f_{\rm app}^{\rm model A2}$ for
$t<4.2$~Myr in the upper panel, suggesting that the initial binary
proportion and period distribution can, in principle, be
constrained. The above inequality is violated for times $t>1$~Myr in
the much smaller sample in the cluster core, and information on the
initial binary proportion is lost.  The model results show that the
observational constraints are still too weak to allow a distinction
between models~A1 and~A2. Both are consistent with the data for
dynamical cluster ages $t\simgreat0.3$~Myr.

The total binary proportion, $f_{\rm tot}$ (equation~\ref{eqn:binf}),
also shown in Fig.~\ref{fig:fappA}, decreases rapidly for
$t\simless0.5$~Myr, and later at a much slower rate. At $t=0$, $f_{\rm
tot}$ is lower than the assumed primordial proportions because wide
binary systems are disrupted through superposition in the crowded
inner cluster region (see Section.~\ref{sec:prbins}). That $f_{\rm
app}$ suffers from substantial chance-projection pairs is evident by
$f_{\rm tot}$ decreasing at a faster rate during $t\simless1$~Myr.

Binary activity leads to an overall expansion of the cluster and
reduction in ambient temperature, i.e. $\sigma_{\rm op,c}$. The
diminished density and velocity dispersion reduce the encounter rate
and rate of binary destruction. Effectively, the binary hardness
boundary shifts to longer periods, making further binary-system
ionising events less probable because the period distribution has been
truncated below the shifting boundary, as discussed in the previous
subsection.  At $t\approx5$~Myr, $f_{\rm tot}^{\rm A1}=0.34$ and
$f_{\rm tot}^{\rm A2}=0.30$, and the surviving binary population is
sufficiently hard that further decrease will not be significant.  In
the cluster, the binary proportion will increase during the subsequent
cluster life-time owing to mass segregation and loss of preferentially
single stars (fig.~3 in Kroupa 1995c).

\begin{figure}
\plotfiddle{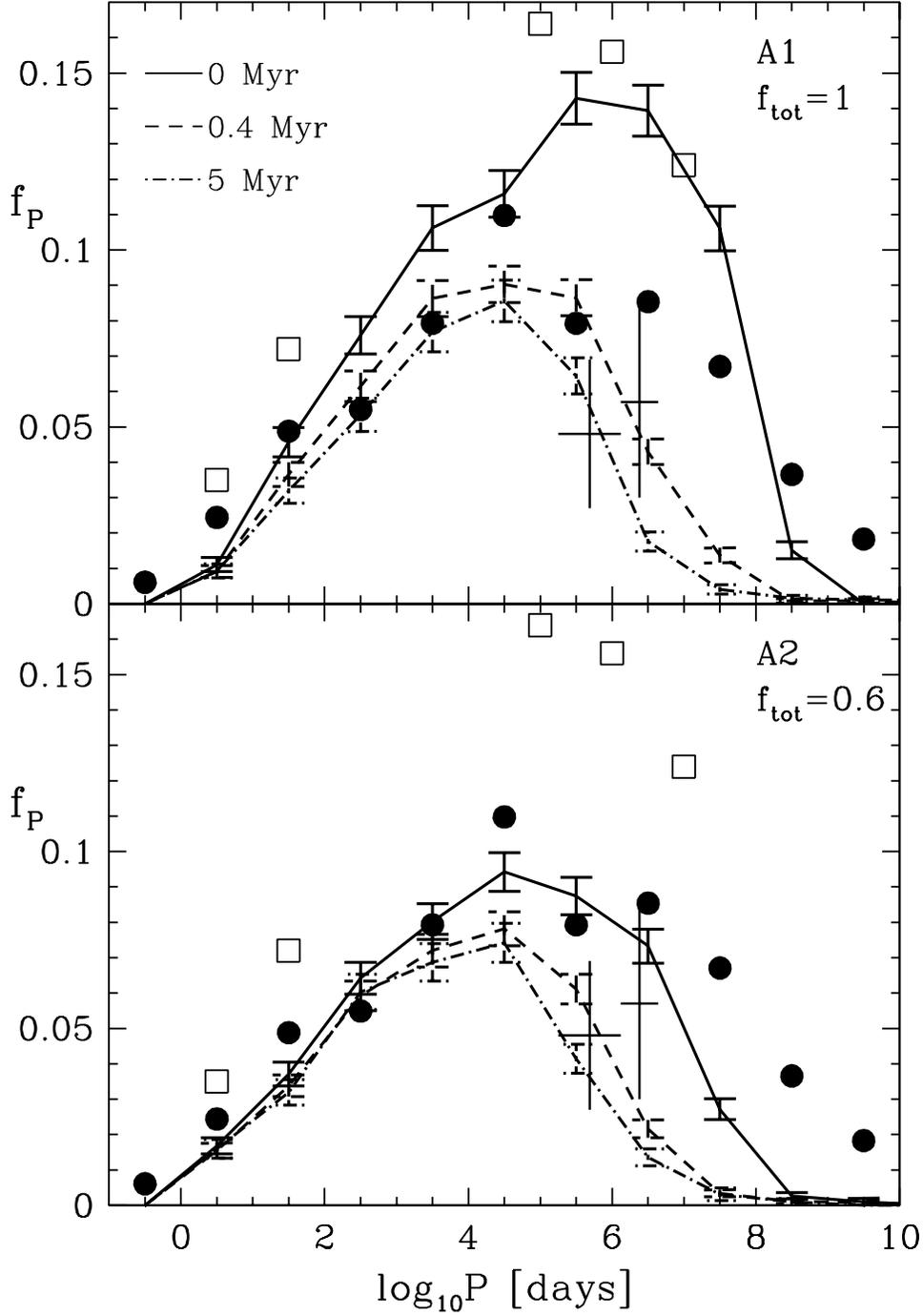}{17cm}{0}{70}{70}{-225}{-15}
\caption{Period distribution for all binary systems within $R\le1$~pc
of the density maximum (higher order systems are not counted).  Upper
panel and lower panels are for models~A1 and~A2,
respectively. Observational data in both panels are as in
Fig.~\ref{fig:init_P}. The two large crosses at log$_{10}P=5.69$ and
log$_{10}P=6.38$ are one-sigma observational constraints for
$r<0.3$~pc and $0.07<r<0.3$~pc, respectively (Petr 1998).
\label{fig:per1}}
\end{figure}

During the first 2~Myr, long-period binary systems are ionised,
leading to a period distribution, $f_{\rm P}$, that is truncated at
progressively shorter periods. Fig.~\ref{fig:per1} demonstrates that
most destruction of long-period ($P>10^6$~days) binaries occurs within
$t\simless0.4$~Myr. There is no significant difference between the
period distributions measured within $R\le0.5$~pc and $R\le1$~pc.
Also, the figure shows that the period distributions in models~A1
and~A2 are similarly truncated by $t\simgreat0.4$~Myr. However, in
model~A1 a somewhat larger proportion of binary systems with $10^5 < P
< 10^7$~days remains than in model~A2. This is a result of the complex
interplay between (i) the cross section for binary-destruction being
proportional to $\sigma^{-2}$ (eqn.~26 in Heggie \& Aarseth 1992) with
the velocity dispersion $\sigma$ being larger in model~A1 for
$1<t<5$~Myr, (ii) the binary destruction rate being proportional to
the square of the density of binaries (eqn.~26 in Heggie \& Aarseth
1992), and (iii) the initially larger binary proportion in model~A1.

\begin{figure}
\plotfiddle{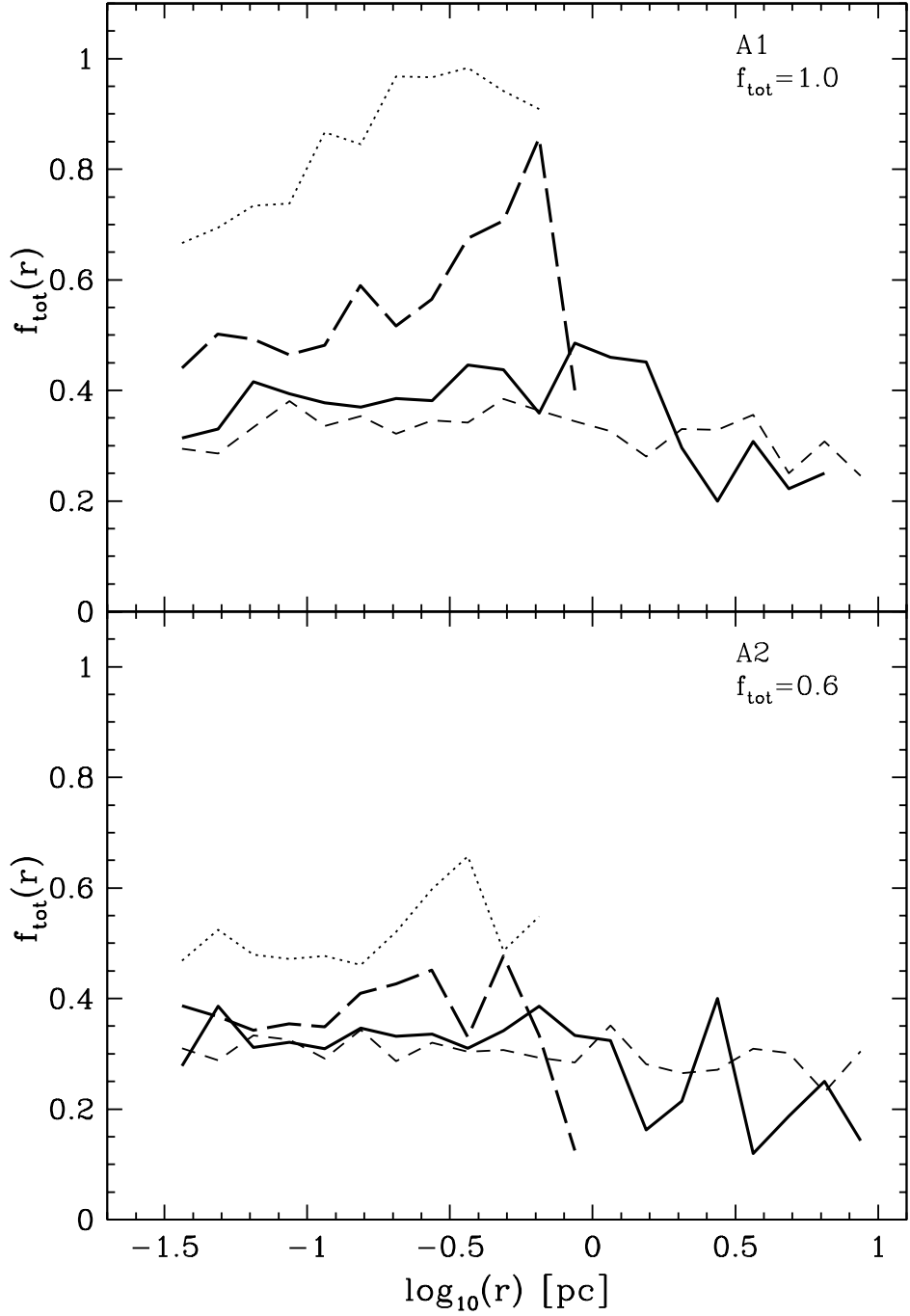}{17cm}{0}{70}{70}{-225}{-15}
\caption{Projected radial dependence of the total binary proportion,
$f_{\rm tot}(r)$ (equation~\ref{eqn:binf}).  In both, the upper
(model~A1) and the lower panel (model~A2), dotted lines are the
initial distributions, long-dashed lines are for $t=0.2$~Myr, solid
lines are for $t=1$~Myr and short-dashed lines are for $t=5$~Myr.
\label{fig:ftot_radA}}
\end{figure}

As is evident from Fig.~\ref{fig:ftot_radA}, initially there are fewer
binaries near the centre in model~A1. This is a result of immediate
disruption of long-period systems from crowding
(Section~\ref{sec:prbins}).  Note that most of the cluster is
initially confined to $r\simless0.6$~pc (log$_{10}r\simless-0.22$), so
that there is no significant difference in the period distributions
sampled within $r\le0.5$~pc and $r\le1$~pc. At $t=0.2$~Myr, the radial
dependence is more pronounced, with binary-star depletion progressing
further out, but the outer regions at $r\approx0.6$~pc retain a high
binary proportion. At this stage of cluster evolution, the period
distribution is already truncated at $P\approx10^6$~days within
$r\simless0.2$~pc (log$_{10}r \simless -0.70$), but remains roughly
primordial at $r\approx0.6$~pc. Further out, however, $f_{\rm tot}$
drops significantly. These are mostly single stars that have been
ejected onto long-period and eccentric orbits in the cluster, and form
a transient binary-deficient young halo population.  After 1~Myr,
there remains a slight radial dependence, with $f_{\rm tot}$
increasing slightly with increasing $r<1.3$~pc (log$_{10}r<0.11$) in
model~A1. Again, the decay for larger $r$ is a result of primarily
low-mass stars being ejected during the disruption of binary systems
near the cluster core.  The radial dependence has vanished by
$t=5$~Myr due to thorough mixing of the stellar population, and
$f_{\rm tot}\approx0.34$.

The lower primordial $f_{\rm tot}=0.6$, together with the Galactic
field Gaussian log-period distribution at birth, leads to no
pronounced radial dependence initially in model~A2, because most
binary systems are too tightly bound to be disrupted through crowding.
During times $t\simless\,{\rm few}\times t_{\rm cross}$,
i.e. $t\simless {\rm few}\times 0.06$~Myr, however, $f_{\rm tot}$ is a
slightly increasing function of $r$, until the entire cluster is
mixed.  A strong decrease in $f_{\rm tot}$ for $r\simgreat0.5$~pc
(log$_{10}r \simgreat -0.30$), and a slight decrease for
$r\simgreat0.6$~pc are evident at $t=0.2$ and 1~Myr, respectively, for
the same reason as above.  However, $f_{\rm tot}\approx0.30$ for
$t>1$~Myr and $r\simless10$~pc, and little further evolution is
apparent until $t=5$~Myr.

Unfortunately the observational constraints cannot ascertain that the
binary proportion is lower in the central parts
(Section~\ref{sec:binf}). The uncertainties remain too large. The
data can be improved by pushing observational resolution to
separations of about 20~AU (0.04~arcsec at a distance of 450~pc) in
order to sample the period distribution around its likely maximum at
log$_{10}P\approx4.5$~[days].  The models show that $f_{\rm tot}$
should be independent of $r$ for $1<t\simless5$~Myr, by which time the
stellar population is well mixed. If observations find evidence for
$f_{\rm tot}(r\approx1.3\,{\rm pc})>f_{\rm tot}(r>1.3\,{\rm pc})$
and/or $f_{\rm tot}(r\approx1.3\,{\rm pc})>f_{\rm tot}(r<0.3\,{\rm
pc})$ then this would be evidence for a Taurus-Auriga-like $f_{\rm P}$
at birth (model~A1), and that the Trapezium Cluster is $t\simless
10\,t_{\rm cross}$ old, remembering that the alternative model~A2 with
a Galactic-field $f_{\rm P}$ does not show significant variations of
$f_{\rm tot}$ with $r$.

\subsection{Expanding clusters -- B}
\label{sec:expand}

\subsubsection{Cluster evolution}
\label{sec:evolB}
\begin{figure}
\plotfiddle{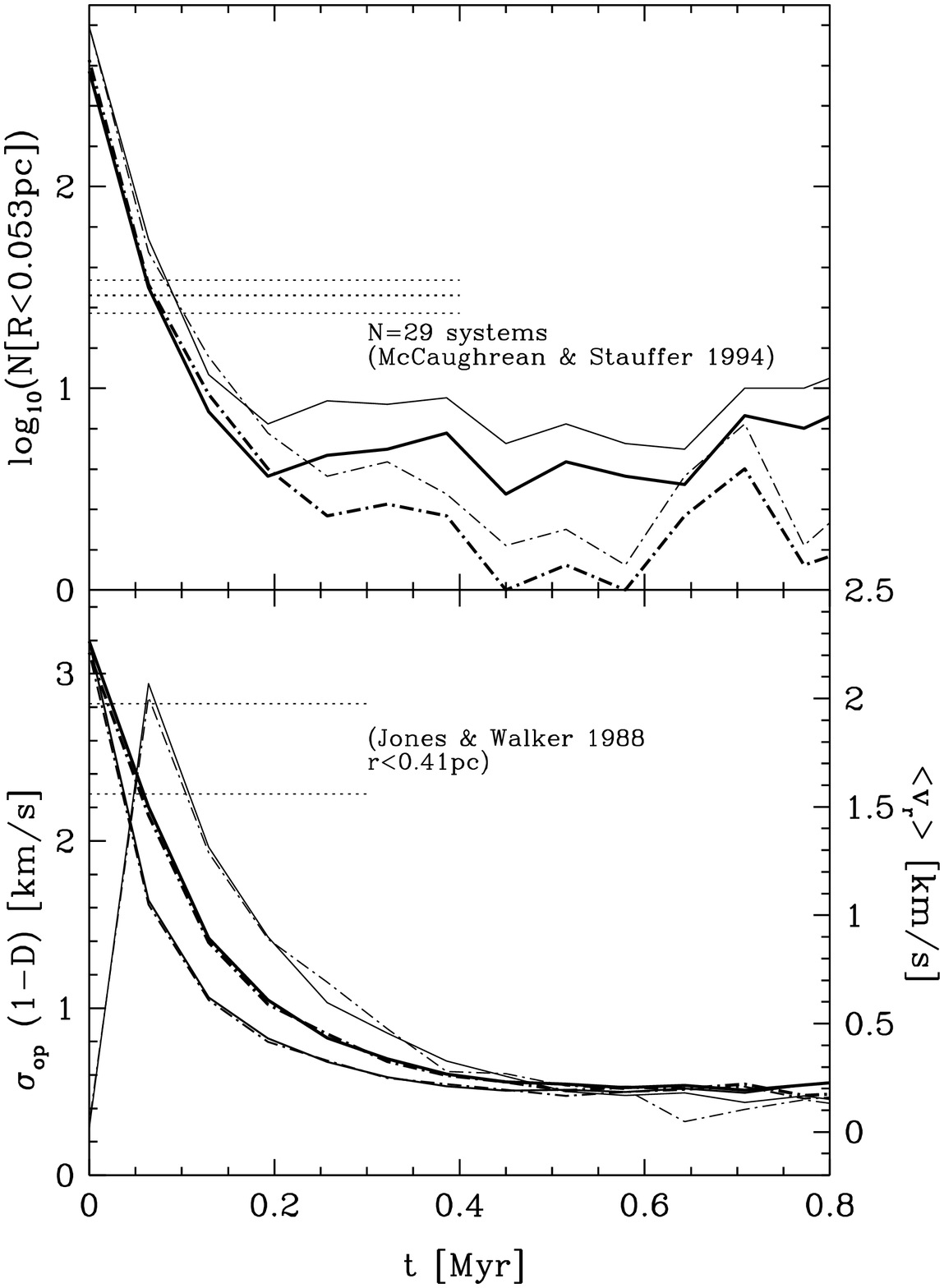}{15cm}{0}{64}{64}{-190}{-20}
\caption{As Fig.~\ref{fig:d_vdA}, but for expanding models~B1 (solid
curves) and~B2 (dot-dashed curves).
\label{fig:d_vdB}}
\end{figure}

\noindent 
The central density is initially as large as in the equilibrium
models, but decreases rapidly (Fig.~\ref{fig:d_vdB}). Noteworthy is
the divergence of the two models (B1 with $f_{\rm tot}=0.77$ and~B2
with $f_{\rm tot}=0.48$ initially, see Fig.~\ref{fig:init_P}) at
$t\approx0.18$~Myr. Thereafter, model~B1, which initially contains
more long-period binary systems, evolves with a constant stellar
number density in the central region, amounting to $N(R<0.053\,{\rm
pc})\approx8$ stars. Again, as in the equilibrium models, the break-up
of long-period binaries causes cluster cooling, which reduces the
velocities and thus aids a part of the expanding cluster to form a
bound object. This point will be investigated in more detail in a
forthcoming paper. In model~B2, which contains fewer long-period
binaries, binary cooling is less effective, and a smaller part of the
cluster condenses out to form a bound entity. In this case,
$N(R<0.053\,{\rm pc})\approx2$ stars, the decay being halted after
$t\approx0.4$~Myr. 

Noteworthy is that the bound clusters that form in models~B1 and~B2
still have central densities $\rho_{\rm c}\approx 10^{4.1}$ and
$10^{3.5}$~stars/pc$^3$, respectively. In comparison, the Pleiades
Cluster has a central density of only about 27~systems/pc$^3$ from
fig.6 in Raboud \& Mermilliod (1998a), or between~14 and~27
systems/pc$^3$ from Pinfield, Jameson \& Hodgkin (1998) if the mean
stellar mass is $1\,M_\odot$ or $0.5\,M_\odot$, respectively.

The projected velocity dispersion within $r<0.41$~pc, which is
initially an order of magnitude larger than in the equilibrium models,
remains equal in both models and decreases rapidly owing to the
expansion against the gravitational potential and the loss of
fast-moving stars from the measurement region ($r\le0.41$~pc).  The
evolution, shown in Fig.~\ref{fig:d_vdB}, slows appreciably after
about 0.2~Myr when only the slow-moving tail of the population remains
within the central radius. Correction for the significant radial bulk
motion is necessary, leading to a reduction in the estimated velocity
dispersion by up to about 0.5~km/s at $t\approx 0.07$~Myr. The initial
radial bulk velocity, $\left<v_{\rm r}\right>=0$, because the velocity
vectors are distributed isotropically. As time progresses, the fast
and slow-moving stars separate out, and a maximum bulk velocity within
$r<0.41$~pc of $\left<v_{\rm r}\right>=2.9$~km/s is reached at
$t\approx0.06$~Myr. The sharp decay thereafter comes from the loss of
the fast-moving stars from the measurement region.

The central number density is consistent at the three-sigma level with
the observational constraint during the time interval $0.05\simless
t\simless 0.1$~Myr, while the velocity dispersion, $\sigma_{\rm
op,c}$, is consistent at the three-sigma level with the observed value
for $0\simless t\simless 0.06$~Myr (Fig.~\ref{fig:d_vdB}).  Thus this
model could be a reasonable description of reality if about 60~per
cent of the mass of the cluster was expelled about 50~thousand years
ago, at which time the Trapezium Cluster would have gone into a rapid
expansion phase.

However, this age is significantly younger than the age inferred from
HR diagram fitting (Section~\ref{sec:numb_age}). If the expanding
model is correct then this age difference can be explained by the
cluster having spent a few~$10^5$~yr in an embedded phase prior to the
postulated gas-expulsion event $\approx5\times10^4$~yr ago.

\subsubsection{The binary stars} 
\label{sec:binariesB}
\begin{figure}
\plotfiddle{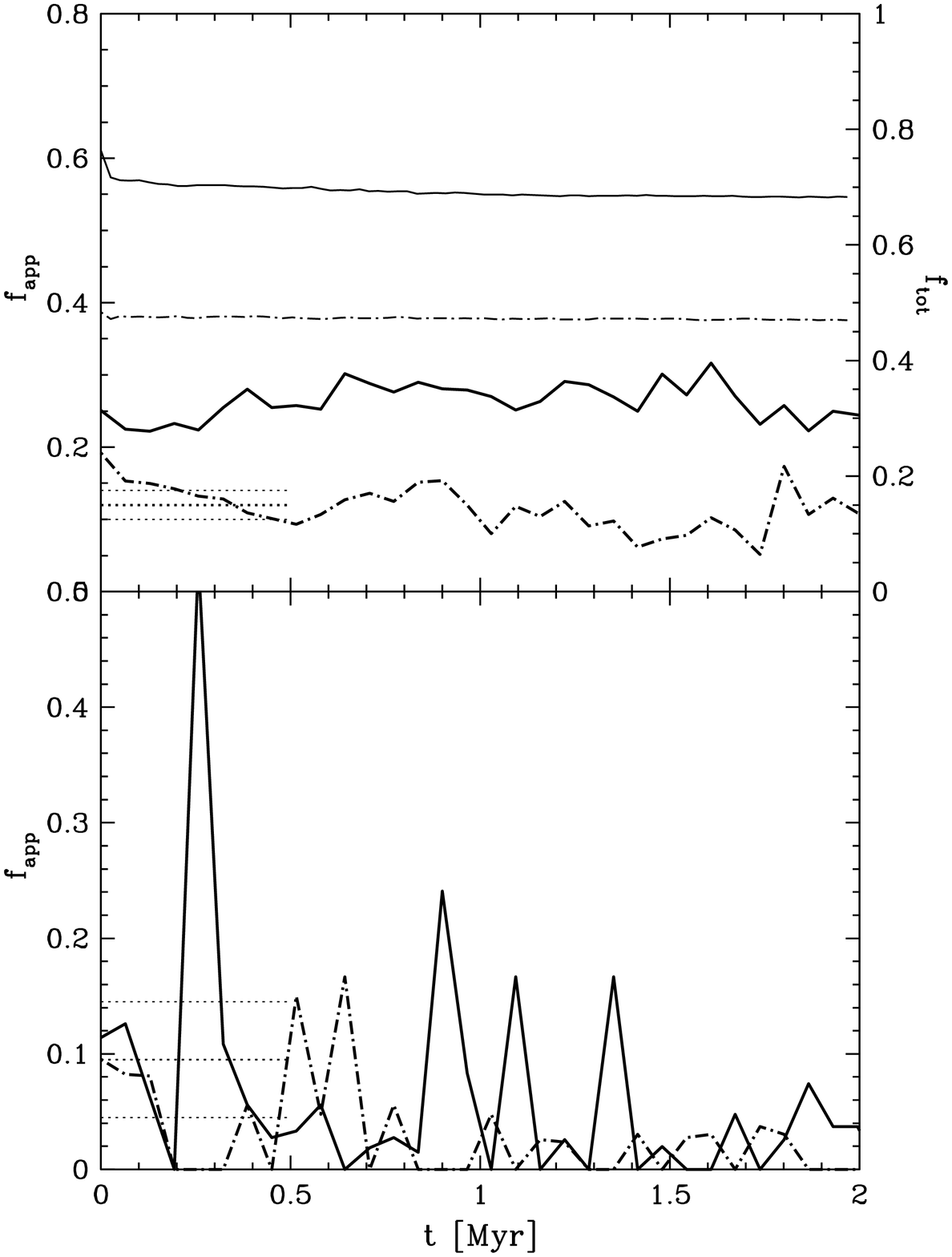}{15cm}{0}{62}{62}{-190}{-20}
\caption{As Fig.~\ref{fig:fappA}, but for expanding models~B1 (solid
curve) and~B2 (dot-dashed curve).
\label{fig:fappB}}
\end{figure}

\begin{figure}
\plotfiddle{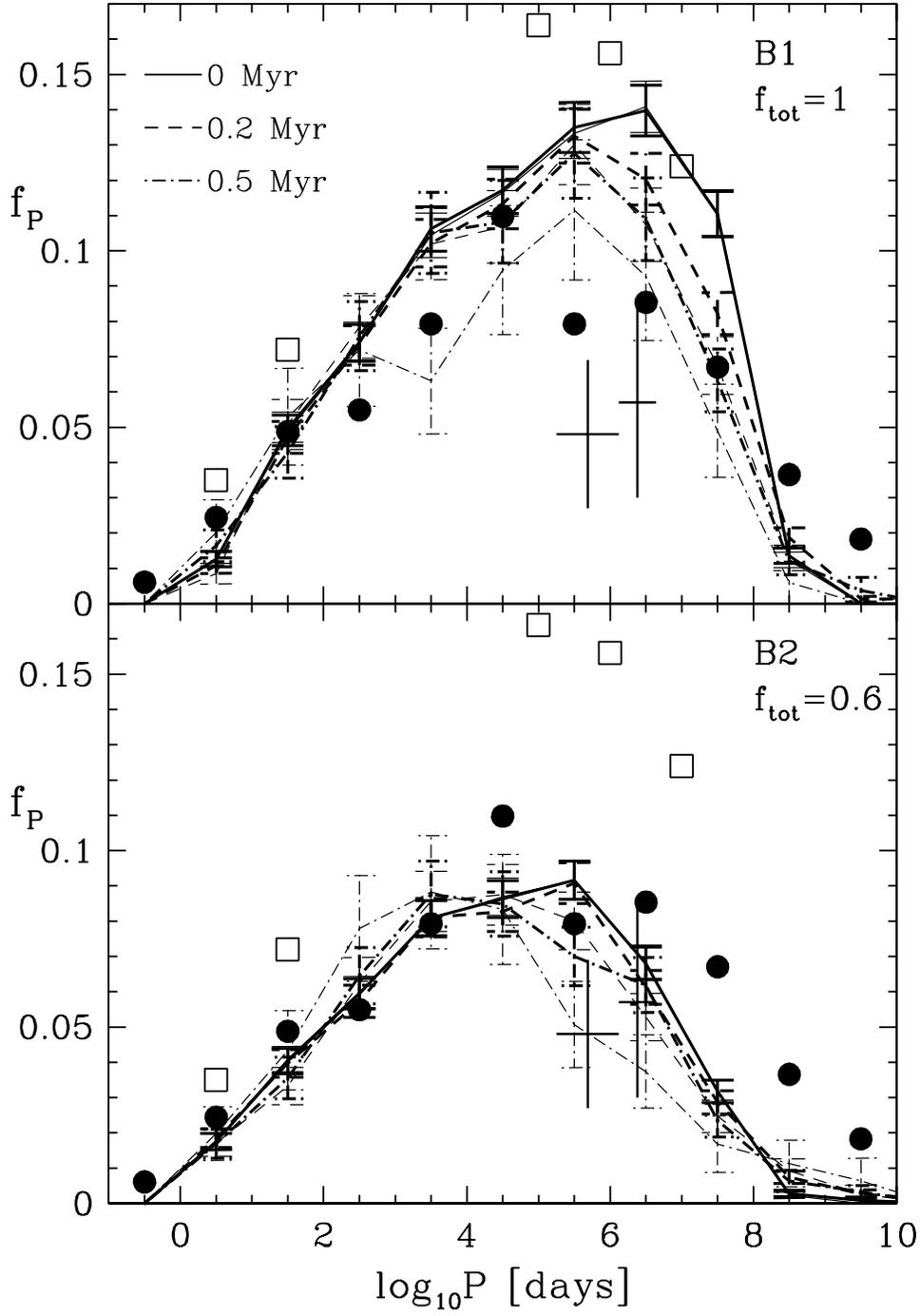}{17cm}{0}{70}{70}{-225}{-15}
\caption{As Fig.~\ref{fig:per1}, but for expanding models~B1 (upper
panel) and~B2 (lower panel). Thick and thin curves are for binaries
with $R\le1$~pc and $R\le0.5$~pc, respectively.
\label{fig:per2}}
\end{figure}

\noindent
Rapid expansion halts the disruption of binary systems at an early
stage. This is evident in Fig.~\ref{fig:fappB}. From this figure it
follows that the model apparent binary proportion is consistent with
the observational constraint at the three-sigma level for
$t\simgreat0.05$~Myr, provided primordial $f_{\rm tot}=0.6$. A smaller
primordial $f_{\rm tot}$ would lead to earlier agreement.  The total
binary proportion shows little evolution after $t\approx0.02$~Myr,
being $f_{\rm tot}=0.68$ in model~B1, and $f_{\rm tot}=0.47$ in
model~B2. The latter value is comparable to the binary proportion of
late-type systems in the Galactic field ($f_{\rm tot}=0.47$, Kroupa
1995a), suggesting that expanding Trapezium-Cluster-type stellar
assemblages could account for the majority of the Galactic field
population.  This is corroborated by the period distribution
(Fig.~\ref{fig:per2}) for log$_{10}P\simless7$. However, the observed
period distribution in the Trapezium Cluster appears to be
significantly below that of the Galactic field for
log$_{10}P\approx7.06 - 8.11$ (Scally, Clarke \& McCaughrean 1999),
which may imply that the Galactic field is not made of unbound
Trapezium-Cluster-type assemblages, unless wide multiple systems form
through capture in the expanding flow. For model~B1, the period
distribution has a surplus of orbits with $10^5<P<10^8$~days.

There is a distinguishable difference between the period distributions
within $R\le0.5$~pc and $R\le1$~pc (Fig.~\ref{fig:per2}). The
distributions undergo no further evolution after $t\approx0.5$~Myr,
but within 0.5~pc, they are somewhat depleted relative to the
distributions obtained for systems within 1~pc.  In particular, Petr's
(1998) observational constraints are in nice agreement with model~B2.
This depletion results from the dynamical interactions within the
remaining low-mass cluster, which has a three-dimensional velocity
dispersion of $\sqrt{3}\,\sigma_{\rm op,c}\approx0.35$~pc/Myr
(Fig.~\ref{fig:d_vdB}). The time-scale for population mixing within
1~pc is thus roughly 3~Myr.

\begin{figure}
\plotfiddle{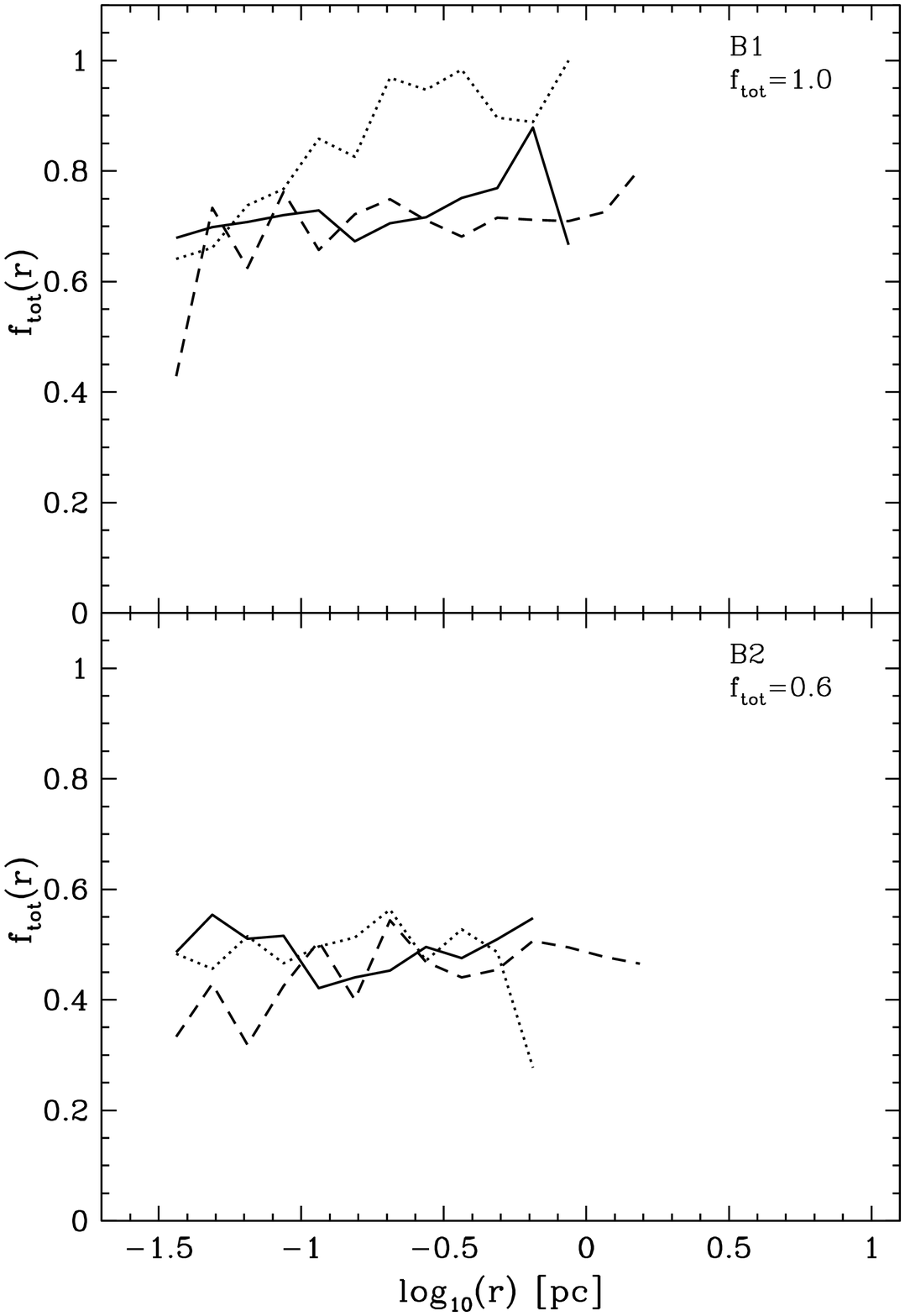}{17cm}{0}{70}{70}{-225}{-15}
\caption{As Fig.~\ref{fig:ftot_radA}, but for expanding models~B1 (upper
panel) and~B2 (lower panel). In both panels, the dotted line is the
initial distribution, and the solid and dashed lines are $f_{\rm
tot}(r)$ at $t=0.064$~Myr and 0.2~Myr, respectively.
\label{fig:ftot_radB}}
\end{figure}

In model~B1, formally with primordial $f_{\rm tot}=1$, the total
binary proportion increases with projected radius initially
(Fig.~\ref{fig:ftot_radB}), which is a result of disruptive crowding
as in model~A1. A slight remaining increase with $r$ is maintained
at later times because the stellar population cannot mix as the
cluster expands. As in model~A2, there is initially no significant
radial dependence of $f_{\rm tot}$ in model~B2.  This is retained at
$t=0.064$~Myr owing to the rapid expansion, and is consistent with the
presently available observational constraints suggesting no
significant radial dependence within $r\approx0.3$~pc
(log$_{10}r\approx-0.52$). At $t=0.2$~Myr, however, both models show a
slightly smaller $f_{\rm tot}$ for $r\simless0.16$~pc
(log$_{10}r\simless-0.80$) than at larger radii, reflecting the
difference in $f_{\rm P}$ for $R\le0.50$~pc and $R\le1$~pc noted
above.

\subsection{Clusters in cold collapse -- C}
\label{sec:coldc}

\subsubsection{Cluster evolution}
\label{sec:evolC}
\begin{figure}
\plotfiddle{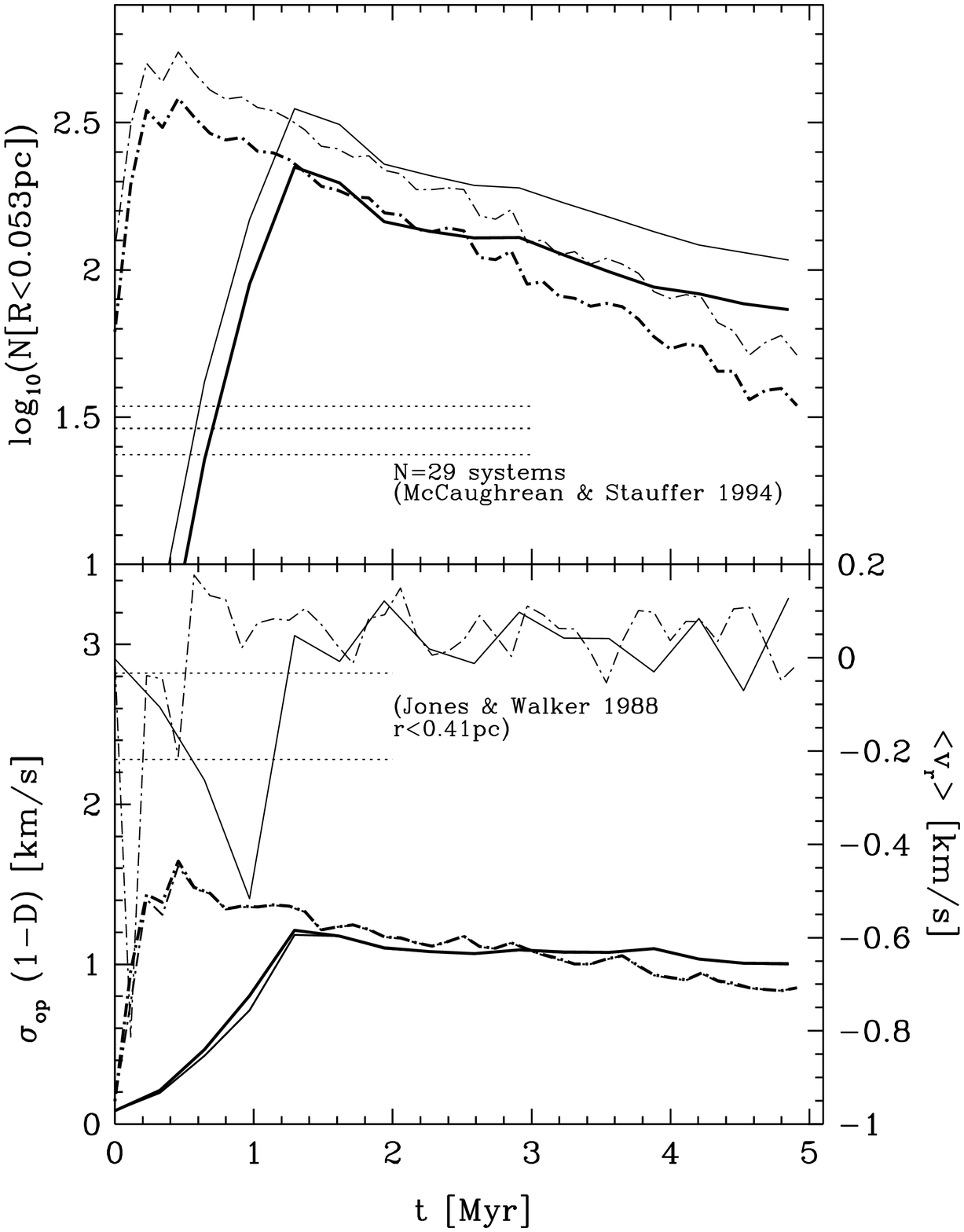}{15cm}{0}{64}{64}{-190}{-20}
\caption{As Fig.~\ref{fig:d_vdA}, but for collapsing models~C1
(dot-dashed curves) and~C2 (solid curves).
\label{fig:d_vdC}}
\end{figure}

\noindent 
Cold collapse from an initial Plummer density distribution leads to a
rapid increase of the central number density, which is maximised at
the cold-collapse (or dynamical) time $t_{\rm cc}\approx0.4$~Myr for
model~C1 ($R_{0.5}=0.4$~pc initially) and at $t_{\rm
cc}\approx1.3$~Myr for model~C2 ($R_{0.5}=0.8$~pc initially).
Model~C2 does not achieve as large a central density as model~C1
because the inner regions virialise before the outer regions have
fallen in.  The evolution is shown in Fig.~\ref{fig:d_vdC}.

The velocity dispersion increases and achieves a maximum at $t_{\rm
cc}$ when the clusters rapidly evolve into new equilibrium states
through violent relaxation (Lynden-Bell 1967). The projected bulk
radial velocity, $\left<v_{\rm r}\right>$, decreases as infall
progresses. For model~C1 the maximum radial bulk motion $\left<v_{\rm
r}\right>=-0.8$~km/s, and for model~C2 it is $-0.5$~km/s
(Fig.~\ref{fig:d_vdC}). The corrections to the projected velocity
dispersion are, however, small.

After virialisation, the clusters evolve on the much longer relaxation
time-scale. It is longer for virialised model~C2, which has a smaller
peak density, and consequently the decay of the central density
proceeds at a slower rate.  In both models, the central density
remains significantly larger within 5~Myr than the observational
constraint, with an exception only for a short moment at
$t\approx0.06$~Myr for model~C2, when the central density briefly
passes through the observational constraint. However, the velocity
dispersion remains significantly smaller than the observational
constraint at all times.

\subsubsection{The binary stars} 
\label{sec:binariesC}
\noindent
Both models~C1 and~C2 are assumed to have, formally, a primordial
binary proportion $f_{\rm tot}=1$. In practice the initial $f_{\rm
tot}$ is slightly smaller than~1 (Fig.~\ref{fig:init_P}) because of
immediate disruption of wide systems through crowding, as in the
models above.

\begin{figure}
\plotfiddle{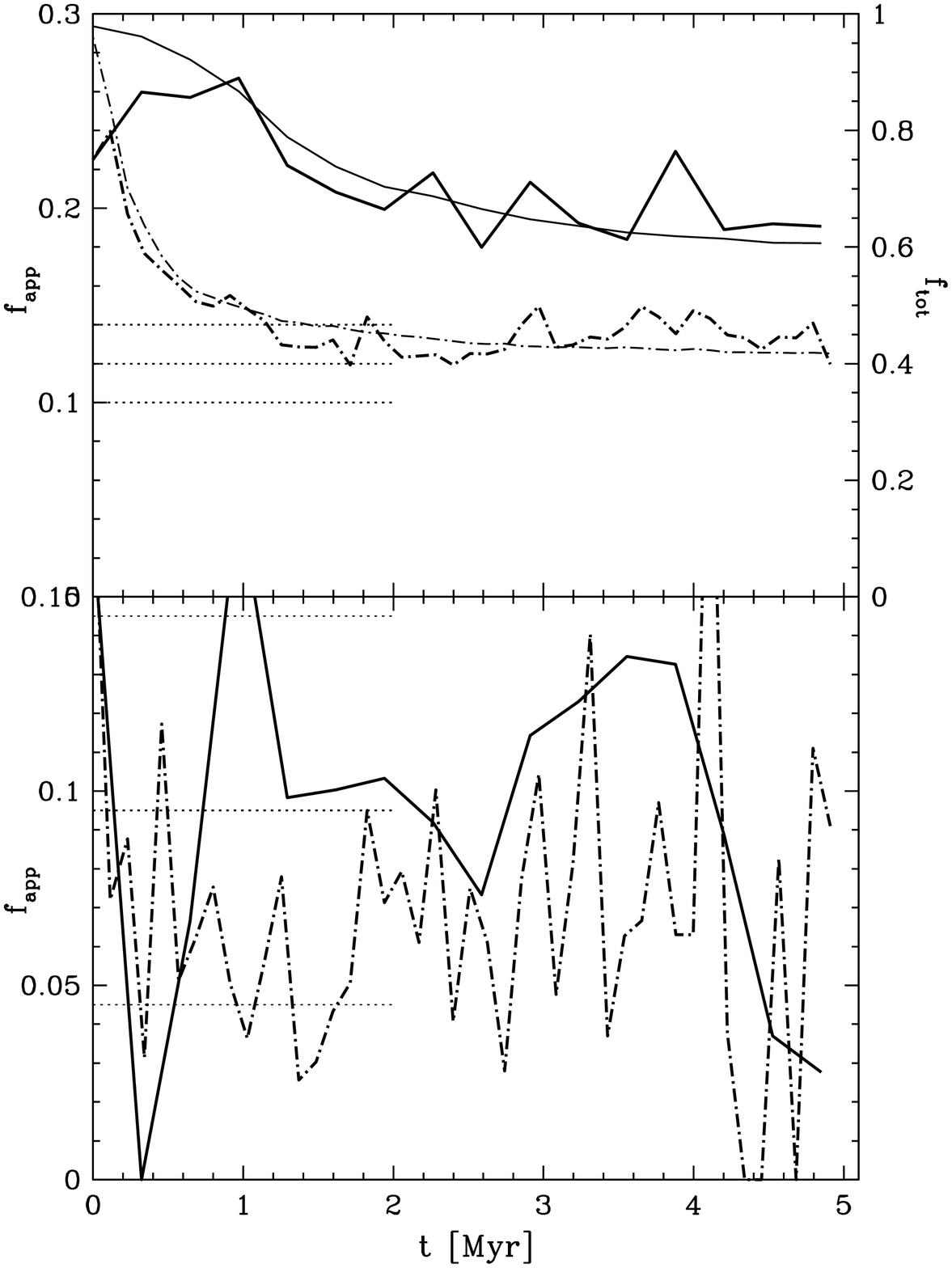}{15cm}{0}{62}{62}{-190}{-20}
\caption{As Fig.~\ref{fig:fappA}, but for collapsing models~C1
(dot-dashed curve) and~C2 (solid curve).
\label{fig:fappC}}
\end{figure}

At $t=t_{\rm cc}$, model~C1 has a central density and velocity
dispersion comparable to the initial ($t=0$) values in model~A1
(Section~\ref{sec:evolA}). The binary population is thus expected to
evolve similarly as in model~A1, keeping in mind that it has already
undergone some evolution during the collapse.  This is confirmed by
comparing Figs.~\ref{fig:fappC} and~\ref{fig:fappA}: the observer
deduces similar apparent binary proportions, with the evolution in
cluster~C1 lagging behind that in cluster~A1 by approximately $t_{\rm
cc}$. The slightly higher apparent and total binary proportion in
model~C1 comes about because the global density remains lower, despite
the central density achieving a value close to the initial central
density in the equilibrium model~A1, and because the velocity
dispersion remains lower (see also the discussion of binary disruption
in Section~\ref{sec:binariesA}).

In model~C2 the density does not increase to as high a value as in
model~C1, so that fewer binary systems are ionised. This leads to an
apparent binary proportion that is larger than the observational
constraints at all times (upper panel in Fig.~\ref{fig:fappC}). During
the initial collapse phase, the apparent binary proportion increases
slightly due to the increased number of chance projection pairs.

\begin{figure}
\plotfiddle{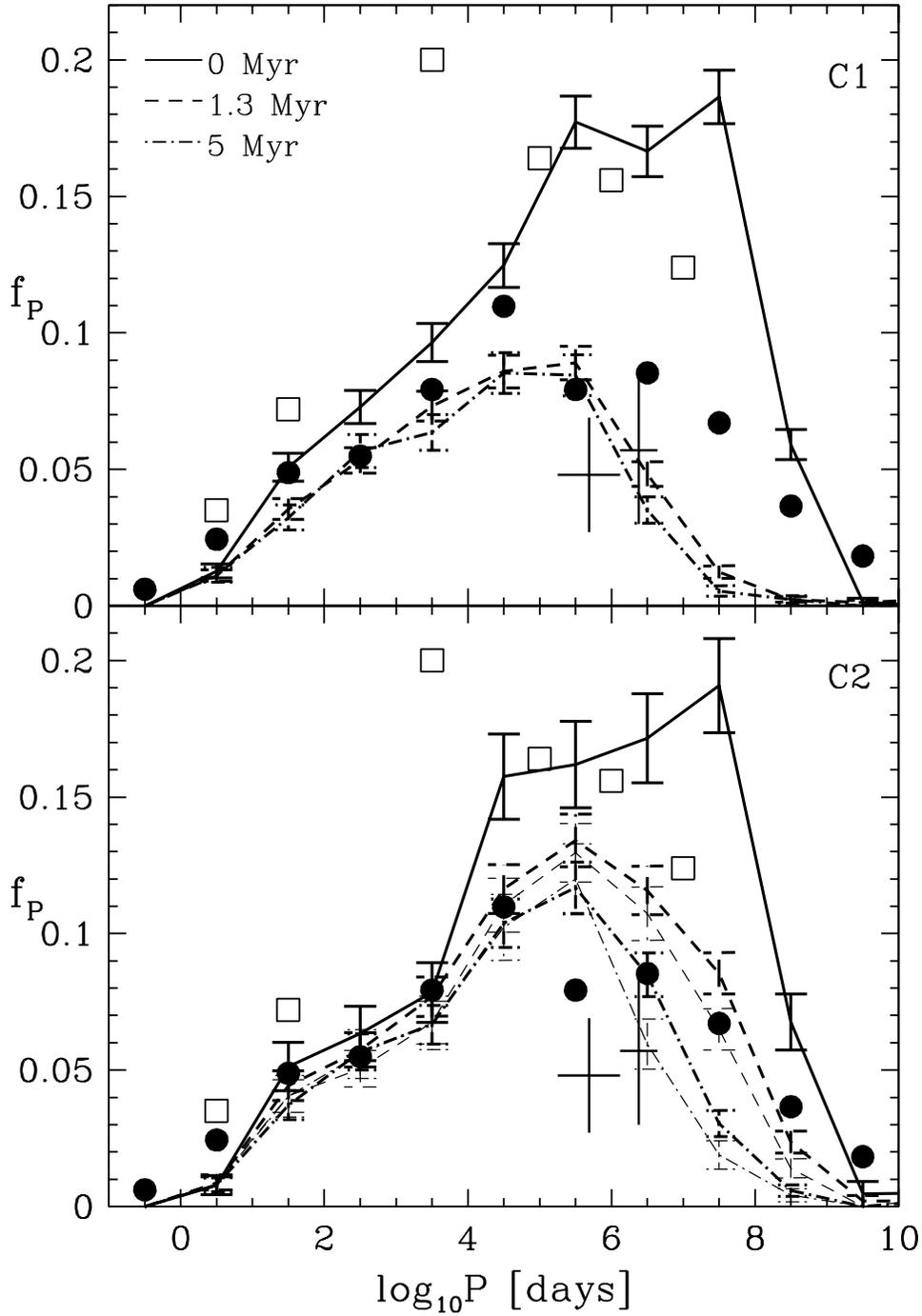}{17cm}{0}{70}{70}{-225}{-15}
\caption{As Fig.~\ref{fig:per1}, but for collapsing models~C1 (upper
panel) and~C2 (lower panel). Thick and thin curves are for $R\le1$~pc
and $R\le0.5$~pc, respectively. The distribution for $R\le0.5$~pc is
only shown in those cases where a significant difference to the
$R\le1$~pc distribution exists.
\label{fig:per3}}
\end{figure}

The period distribution, $f_{\rm P}$, at different times and different
radii in the cluster is shown in Fig.~\ref{fig:per3}. By $t=1.3$~Myr
model~C1 has already virialised and mixed, and there is no subsequent
radial nor any time variation of $f_{\rm P}$. The evolution of $f_{\rm
P}$ through cold collapse is, however, evident in model~C2.  At
$t=1.3$~Myr, the collapse has arrived at the maximum central density,
and $f_{\rm P}$ is already depleted at long periods. There is a larger
depletion of $f_{\rm P}$ in the more concentrated inner region
($R\le0.5$~pc).  At $t=5$~Myr, the cluster is virialised, and $f_{\rm
P}$ is even more depleted at long-periods. The maximum
three-dimensional velocity dispersion is $\sigma=\sqrt{3}\,\sigma_{\rm
op,c}=2.8$~km/s for model~C1, and $\sigma=2.1$~km/s for model~C2
(Fig.~\ref{fig:d_vdC}). Taking these values for the circular orbital
velocity of a binary system with a mass of $1\,M_\odot$, orbital
periods of $10^{5.6}$~days and $10^{6.0}$~days, respectively, are
obtained.  These values reflect the cutoff periods evident in
Fig.~\ref{fig:per3}, and demonstrate the dependence of the rate of
disruption of binary systems on the ambient velocity dispersion in a
self-gravitating stellar system.

\begin{figure}
\plotfiddle{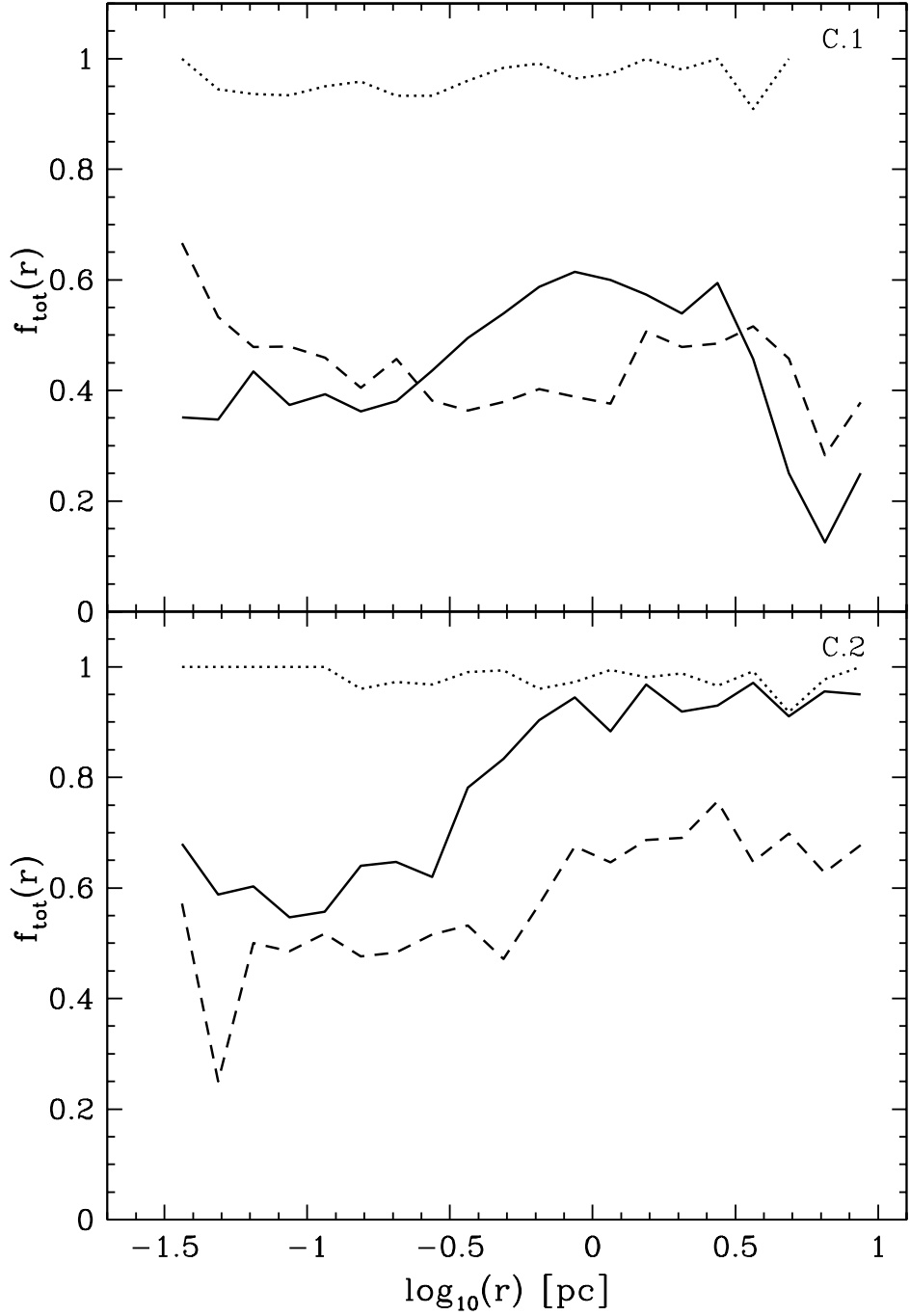}{17cm}{0}{70}{70}{-225}{-15}
\caption{As Fig.~\ref{fig:ftot_radA}, but for collapsing models~C1
(upper panel) and~C2 (lower panel). In both panels, the dotted line is
the initial distribution, and the solid and dashed lines are $f_{\rm
tot}(r)$ at $t=1.3$~Myr and 5~Myr, respectively.
\label{fig:ftot_radC}}
\end{figure}

How cold collapse influences the radial distribution of binary systems
is shown in Fig.~\ref{fig:ftot_radC}. Initially $f_{\rm tot}\approx1$
for all $r$.  After one overall free-fall time, $t=t_{\rm
cc}\approx1.3$~Myr for model~C2, the binary proportion is reduced to
60~per cent within $r\approx0.25$~pc (log$_{10}r\approx -0.60$), where
most of the action during virialisation occurs. At larger radii,
$f_{\rm tot}$ increases with $r$ and remains essentially unaffected
for $r>1$~pc, i.e. in regions of the cluster that have not yet fallen
in. At $t=5$~Myr, however, the entire cluster has virialised, and the
binary proportion is reduced at all radii, with a remaining slight
increase towards the outermost less dense regions of the cluster
reflecting some remaining incomplete mixing of the population. In
model~C1 the collapse time-scale is much shorter ($\approx0.4$~Myr),
even for the outermost radii, so that the binary proportion is reduced
at all radii by $t=1.3$~Myr. At this time, $f_{\rm tot}$ increases
slightly with $r$ for $r\simless0.8$~pc (log$_{10}r\simless -0.10$).
This comes about because the cluster is still not completely mixed.

During the violent relaxation phase binary systems are ionised at a
larger rate owing to the increased number density. Some of the
liberated low-mass stars are ejected to larger radii, leading to the
significant drop in $f_{\rm tot}$ at $r\simgreat2.5$~pc
(log$_{10}r\simgreat 0.40$, compare with model~A1 in
Fig.~\ref{fig:ftot_radA}). Such a decrease is not evident very well in
model~C2 because fewer binary systems are ionised in the
less-concentrated core, providing fewer single stars to the outer
cluster radii. A slight drop is, however, visible at $r>3$~pc
(log$_{10}r>0.48$) in model~C2.

By $t=5$~Myr the binary proportion is larger in the centre of model~C1
than at larger radii. This is a result of dynamical mass segregation,
and is observed in other young clusters (e.g. Elson et al. 1998;
Raboud \& Mermilliod 1998b).

\section{DISCUSSION}
\label{sec:disc}
\noindent
Six different models of the Trapezium Cluster, summarised in
Table~\ref{table:models}, are evolved using Aarseth's {\sc Nbody5}
code. They are clusters in virial equilibrium, expanding clusters
assuming a star formation efficiency of 42~per cent and instant gas
expulsion, and clusters in cold collapse, assuming the stars freeze
out of the molecular cloud with a small relative velocity dispersion.
We confront the evolving model population with observational
constraints to infer if and how binary properties vary with
star-forming conditions, and if the present and past dynamical state
of the Trapezium Cluster can be inferred.

The virial equilibrium (A1 and A2) and initially collapsing (C1 and
C2) models have, in comparison with the Trapezium Cluster, at all
times too high a central number density, while simultaneously having
too small a velocity dispersion. These are thus excluded as realistic
stellar-dynamical models of the Trapezium Cluster. However, we find
that one of the expanding models is a solution that meets the
observational constraints (Section~\ref{sec:youngcl} below).

\subsection{Variation of binary proportion with radius}
\label{sec:fr}
\noindent
All models show that the variation of the binary proportion with
radius, $f_{\rm tot}(r,t>0)$, is a diagnostic of the dynamical state
of a cluster (Figs.~\ref{fig:ftot_radA}, \ref{fig:ftot_radB}
and~\ref{fig:ftot_radC}).

For example, if the binary proportion near the centre of the cluster
is lower than in its outer regions, then this indicates that the
cluster is younger than a few crossing times. If the binary proportion
is constant throughout most of the cluster but lower in its outermost
regions, then this implies that the cluster is old enough for the
stellar population to be well mixed, but young enough for binary
activity in the inner regions to have ejected single stars into a
halo, i.e. it is a few to ten crossing times old. If the binary
proportion near the cluster centre is larger than at larger radii,
then this indicates that the cluster is sufficiently old for dynamical
mass segregation to have had time to occur.

In particular, the collapsing models demonstrate how
binary-destruction propagates to larger radii as collapse progresses,
with a well-developed radial dependence of $f_{\rm tot}$ that is to a
large extend eradicated after a few crossing times through mixing of
the stellar population.

\subsection{Evolution of the binary proportion}
\label{sec:ft}
\noindent
Despite not being possible solutions, the equilibrium and collapsing
models shed important insights on the evolution of the binary
population during the first few~Myr of a cluster's lifetime after gas
dispersal (Figs.~\ref{fig:fappA}, \ref{fig:fappB}
and~\ref{fig:fappC}).

The calculations show that binary destruction is very efficient in a
dense star cluster. The binary proportion decreases from its initial
high value to less than 40~per cent within 1--2~Myr in our virial
equilibrium cluster model, with most destruction happening within the
first 0.5~Myr. Ultimately, expansion of the cluster shifts the
hard/soft binary boundary to longer periods that have already been
depleted, thus halting further reduction of the binary population, and
reducing the cluster expansion rate. An equilibrium is reached between
the global properties of the cluster and its remaining binary
population after about~2~Myr.

This implies that the distribution of binary-star periods, in
particular the truncation period, in a Galactic cluster is a measure
of its dynamical state at birth, or at the moment of maximum
compression (e.g. after a cold collapse).

\subsection{Binary cooling}
\label{sec:bc}
\noindent
The effects of binary activity on the evolution of the central density
and velocity dispersion are clearly noticeable in the equilibrium
models (Fig.~\ref{fig:d_vdA}).

A large proportion of binaries near the hard/soft interface
immediately causes overall cluster expansion, because their activity
liberates kinetic energy.  This stands in contrast to the immediate
onset of core contraction for a single-star cluster.

However, an initially larger proportion of soft binaries (model~A1)
leads to a cooling effect, because kinetic energy is used up in
ionising these binary systems. As a result, the cluster expands at a
slower rate, and the central number density remains higher than in
model~A2 in which there are fewer such primordial binaries.  The ratio
of the number of binaries with periods near the hard/soft boundary to
the number of soft binaries is thus a critical quantity driving the
initial rate of expansion.

\subsection{A 50\,000~yr old Trapezium Cluster?}
\label{sec:youngcl}
\noindent
The expanding models (B1 and B2) are consistent with the observational
constraints, provided expansion began $t_{\rm
exp}\approx5\times10^4$~yr ago. This might be possible if about 60~per
cent of the mass in the cluster was expelled at that time.  Owing to
the rapid expansion of models~B1 and~B2, the initial binary population
remains essentially unchanged.  

In these models, the binary proportion at a time $t_{\rm exp}$ ago
must have been $f_{\rm tot}\approx0.48$ to be consistent with the
observational constraints. This is lower than the binary proportion in
Taurus--Auriga, and, if model~B2 does represent reality, may be due to
dynamical evolution in the embedded cluster prior to gas expulsion, or
due to a dependency on cloud temperature as suggested by Durisen \&
Sterzik (1994). This would support the assertion that most
Galactic-field stars stem from Trapezium-Cluster like assemblages,
provided the discrepancy in the proportion of wide binaries in the
Trapezium Cluster and the Galactic field can be accounted for
(Section~\ref{sec:binariesB}).

The finding that expanding model~B2 is consistent with the
observational constraints suggests that a gas mass of approximately
$1000\,M_\odot$ may have been expelled about $5\times10^4$~yr
ago. This quantity of ejected gas is consistent with observations of
outflows from young OB stars (Churchwell 1997). The observations show
that hundreds of $M_\odot$ of gas are expelled within a few to tens of
$10^4$~yr per massive star.  Our assumption of instantaneous mass loss
is thus reasonable. A neutral gas lid lies in front of the Trapezium
Cluster as viewed from Earth, but material is streaming out of the
blister SW of $\theta^1$~Ori~C, where lid extinction is low and
vanishes (O'Dell \& Wen 1992; Wilson et al. 1997; O'Dell 1999). It is
feasible that the massive outflows from the OB stars, that must have
existed in the past, escaped from the cluster in that direction.

The possible rather short expansion age of the Trapezium Cluster is
interesting in the context of the existence of circum-stellar material
around many of the low-mass stars in the Trapezium Cluster.  The
irradiation by the most massive star in the cluster, $\theta^1$~Ori~C,
poses a very destructive environment for these objects.  Bally et
al. (1998) model their photo-evaporation, and find that the ablation
is so rapid, that the photo-ionization age of the blister in which the
Trapezium Cluster is located may be as short as $10^4$~yr. If the age
were a few $10^5$~yr, most of the circum-stellar material seen around
the majority of low-mass stars should have been removed.  O'Dell
(1999), however, points out that the circum-stellar material may have
quasi-equilibrium atmospheres with a much lower mass-loss rate,
implying that these objects could be as old as HR diagram fitting
implies the cluster to be (Section~\ref{sec:numb_age}). 

At present the issue is unsettled, and the present stellar-dynamical
models add tantalising evidence in favour of the Trapezium Cluster
becoming visible only a few $10^4$~yr ago by $\theta^1$~Ori~C
``turning on''.

\subsection{Caveats}
\label{sec:cavs}
\noindent
The main shortcoming of the present study is that the Trapezium
Cluster is treated as an isolated entity. In reality, it appears to be
the core of the much more massive and extended Orion Nebula Cluster
(ONC).  Also, the ONC is partially embedded in the parent elongated
molecular gas cloud, the remaining parts of which lie at least about
0.2~pc behind the most massive Trapezium Cluster star,
$\theta^1$~Ori~C (Wen \& O'Dell 1995).  Hillenbrand \& Hartmann (1998)
estimate that the mass in the stellar component of the ONC and in the
gas may together suffice to produce the observed stellar velocity
dispersion. 

It is clear that future models will need to include the ONC, and at a
later stage the molecular cloud, being computationally much more
expensive. However, even in this context and given the present
results, it can be envisioned that the dense core of the ONC, namely
the Trapezium Cluster, may have begun expanding a few~$10^4$~yr ago.

A companion paper (Kroupa 1999) generalises the present results to
different values of $N$ and $R_{0.5}$, thereby identifying the
candidate sets of models consistent with the observational constraints,
if the ONC is in virial equilibrium, expanding or undergoing a cold
collapse.

Finally, the results presented here assume the clusters are initially
Plummer density distributions. This is realistic in the sense that
star-forming cores are centrally concentrated, which has been used in
the past to argue that Plummer distributions are reasonable initial
conditions for young star clusters (Lada, Margulis \& Dearborn
1984). We expect all density distributions that are centrally peaked
and that have the same central densities, masses and half-mass radii
as the models here to lead to a very similar evolution of the binary
population, although the detailed radial dependence at some time $t$
may be slightly different. The greatest differences are expected for
clusters that initially consist of groupings of aggregates, in the
sense of a hierarchical structure. N-body calculations of the
evolution of a cluster of sub-clusters, each containing approximately
a dozen binary systems, have to be performed in order to address this
particular set of possible models of the Trapezium Cluster and the
ONC. However, here again we expect similar evolution in the final
bound cluster after the sub-clusters have merged, although the number
of escaping stars will be higher during the first few global crossing
times (Aarseth \& Hills 1972).

\section{CONCLUSIONS}
\label{sec:conc}
\noindent
This paper presents fully self-consistent stellar-dynamical models of
the evolution of the Trapezium Cluster over a time-span of 5~Myr,
assuming it consists of 1600~stars with a mass of $700~\,M_\odot$. The
calculations treat model clusters that are (i) in virial equilibrium,
(ii) expanding and (iii) undergoing cold collapse. All models have a
very large primordial binary population.  These models constitute the
hitherto most realistic stellar-dynamical calculations of a young
cluster in existence, covering the time after gas expulsion but
before stellar evolution becomes significant. Thus, cooling of a
cluster through the ionization of long-period binaries has been
discovered for the first time to be a noticeable effect, if the
primordial period distribution is similar to what is seen in the
Taurus--Auriga star formation complex.

The calculations show that a young compact cluster, that is similar to
the Trapezium Cluster and in virial equilibrium, undergoes significant
dynamical evolution during the first few~Myr. The same holds true for
its binary population, which is depleted rapidly, but then stabilises
at $f_{\rm tot}\approx0.30, 0.34$ if the initial period distribution
is as in the Galactic field or Taurus--Auriga-like, respectively.
This work also demonstrates which observational constraints are
critical for discerning between models of very young clusters.  These
are the inner and outer binary proportions, and the radial bulk
motion.

In a young cluster the period distribution is truncated at a period
that depends on the initial dynamical state of the cluster, with
little further evolution after a few~Myr, or after violent relaxation
has occurred. Measurement of the period distribution, or at least of
the truncation period, in a cluster such as the Pleiades or Hyades,
thus constrains the possible range of initial configurations of these
clusters.

We are particularly interested to have discovered that the expanding
cluster model is in best agreement with the observational
constraints. This model assumes that about $1000\,M_\odot$ of gas was
expelled from the cluster about $50\,000$~yr ago, implying that the
binary proportion at that time must have been significantly below the
Taurus--Auriga value, and that the Trapezium Cluster is unlikely to
form a long-lived bound Galactic cluster.  The expansion rate is thus
a very potent observable that urgently needs to be determined. The
present models have a projected bulk radial velocity within a central
radius of 0.41~pc of about 2~km/s at an expansion age of 50\,000~yr.

While the present models are not a unique set of stellar-dynamical
descriptions of the Trapezium Cluster, the present results will form
an incentive for future observational studies (i) of the binary
proportion in the inner and outer regions of very young star clusters,
and (ii) of the velocity dispersion and bulk radial motions in the
clusters. The results, and in particular their generalisation to a
larger range of parameter space, will also be a strong incentive for
conducting further, much more CPU-intensive numerical experiments,
which are already being planned and partially underway.

\acknowledgements 
\vskip 10mm
\noindent{\bf Acknowledgements}
\vskip 3mm
\noindent
We are very grateful to Sverre Aarseth for freely providing {\sc
Nbody5}, and for his unfailing, friendly and uniquely efficient
``customer support''.  We are also thankful to Bob O'Dell for very
useful correspondence, to Tom Megeath and Phil Myers for interesting
conversations, and Gijs Nelemans for helpful comments.  PK thanks Ron
Burman and other staff of the Department of Physics at the University
of Western Australia, where part of this paper was written, for their
kind hospitality during March~1998, and the staff of the
Harvard-Smithsonian Centre for Astrophysics for a very pleasant visit
during March--June~1999, where this paper was finished.  PK
acknowledges support from DFG grant KR1635.

%

\clearpage

\begin{table}
{\small
\begin{minipage}[t]{16cm}
\hspace{3cm}
  \begin{tabular}{*{8}{c}{l}}
   \tableline\tableline
    1 &2 &3 &4 &5 &6 &7 &8 &9\\
      Model  &$R_{0.5}$  
           &$N_{\rm c,sys}$ &$N_{\rm c,st}$ &log$_{10}(\rho_{\rm c})$
           &$Q$ &$f_{\rm tot}$ 
           &log$_{10}P_{\rm max}$ &period distribution \\
           &{\scriptsize (pc)} & & &(pc$^{-3}$) 
           & & &(days)\\
    \tableline\tableline
    A1 &0.10 &376 &624  &5.92    &0.50 &1.0  &8.43 &K2\\
    A2 &0.10 &441 &624  &5.92    &0.50 &0.6  &11.0 &DM91\\
    B1 &0.10 &376 &624  &5.92    &1.20 &1.0  &8.43 &K2\\
    B2 &0.10 &441 &624  &5.92    &1.20 &0.6  &11.0 &DM91\\
    C1 &0.40 &60  &120  &4.12    &0.01 &1.0  &8.43 &K2\\  
    C2 &0.80 &3   &6    &3.21    &0.01 &1.0  &8.43 &K2\\
    \tableline\tableline
\end{tabular}
\end{minipage}
}
\caption{Initial conditions for Trapezium Cluster models. Three
calculations are performed for each model. K2 is Kroupa (1995b), and
DM91 is Duquennoy \& Mayor (1991)}
\label{table:models}
\end{table}
\clearpage
\newpage

\clearpage
\newpage 

\clearpage
\newpage 

\clearpage
\newpage 

\clearpage
\newpage 

\clearpage
\newpage 

\clearpage
\newpage 

\clearpage
\newpage 

\clearpage
\newpage 

\clearpage
\newpage 

\clearpage
\newpage 

\clearpage
\newpage 

\clearpage
\newpage 

\end{document}